\documentclass[twocolumn,aip,jmp,amscd]{revtex4}
\usepackage{amssymb}
\usepackage{amsmath}
\usepackage{amsthm}
\usepackage{graphicx}

\begin{document}

\title{{\itshape Combinations of coupled cluster,
density functionals, and the random phase 
approximation for describing static and dynamic correlation, 
and van der Waals interactions}}

\author{
  Alejandro J. Garza,$^{a}$
  Ireneusz W. Bulik,$^{a}$
  Ana G. Sousa Alencar,$^{a}$
  Jianwei Sun,$^{b}$ John P. Perdew,$^{b,c}$ and
  Gustavo E. Scuseria$^{a,d}$$^{\ast}$\thanks{$^\ast$Corresponding author. Email: guscus@rice.edu
\vspace{6pt}}
\\\vspace{6pt}  $^{a}${\em{Department of Chemistry, Rice University, 
  Houston, Texas 77251-1892, USA}}; 
$^{b}${\em{Department of Physics, Temple University, Philadelphia, 
    PA 19122}}; 
$^{c}${\em{Department of Chemistry, Temple University, Philadelphia, 
    PA 19122}}; 
$^{d}${\em{Department of Physics and Astronomy, Rice University,
  Houston, Texas 77251-1892, USA}}
}

\vspace{6pt}

\begin{abstract}
Contrary to standard coupled cluster doubles (CCD) and 
Brueckner doubles (BD), singlet-paired analogues of
CCD and BD (denoted here
as CCD0 and BD0) do not break down when static correlation is 
present, but neglect substantial amounts of dynamic correlation.  
In fact, CCD0 and BD0 do not account for 
any contributions from multielectron excitations involving only same-spin 
electrons at all. 
We exploit this feature to add---without introducing double 
counting, self-interaction, or increase in cost---the missing correlation 
to these methods via meta-GGA density functionals (TPSS and SCAN).
Furthermore, we improve upon these CCD0+DFT blends by 
invoking range separation: the short- and long-range
correlations absent in CCD0/BD0 are evaluated with DFT and the 
direct random phase approximation (dRPA), respectively. 
This 
corrects the description of long-range van der Waals forces. 
Comprehensive benchmarking shows that the combinations 
presented here are very accurate for 
weakly correlated systems, while also
providing a reasonable description of strongly correlated
problems without resorting to symmetry breaking. 

\end{abstract}

\maketitle


\section{Introduction}

Although coupled cluster (CC) theory is one of the pillars of 
quantum chemistry, commonly used CC methods (\textit{e.g.}, singles 
and doubles, CCSD) often fail in the presence of static 
(or strong) correlation~\cite{Fan2006,Bulik2015}. 
These failures are connected to instabilities that appear in
the random phase approximation (RPA) when Hartree--Fock (HF) 
becomes unstable towards
symmetry breaking~\cite{Bulik2015}---RPA-like terms are contained in the 
equations of traditional coupled cluster 
techniques~\cite{Scuseria2008,Scuseria2013,Peng2013}.
(RPA instabilities come in two flavors: ph-ph and pp-hh. The
former is related to spin and the latter to number symmetry
breaking. ``Instability'' in the CC context means that the solution to
the CC equations results in large or complex cluster
amplitudes and unbound correlation energies.)
Such instabilities may be avoided by modifying the cluster operator 
to include only singlet-paired operators as in, \textit{e.g.}, 
pair CCD~\cite{Limacher2013,Limacher2014,Tecmer2014,Boguslawski2014,%
Stein2014,Henderson2014}
(pCCD) and the recently proposed CCD0 method~\cite{Bulik2015}.
However, this workaround is not a panacea: the suppression of 
terms in the cluster operator inevitably leads to neglecting
part of the correlation. 
For example, CCD0 contains none of the correlation terms
arising from excitations involving same-spin electron pairs 
which are present in CCD (see below).
Despite this clear problem, 
pCCD and CCD0 are proper wavefunction theories: 
both the \textit{ansatz} and its solution are rigorous approximations 
to solving Schr\"{o}dinger equation.
The main objective of the present work is to 
investigate ways to add the correlation that is missing in CCD0 
via physically-motivated combinations of
this method, and its extension using Brueckner orbitals 
(denoted as BD0), with techniques based 
on density functional theory (DFT) and the RPA.

The idea of merging techniques like
CCD0 with DFT
is inspired by a large number of 
previous works~\cite{Lie1974a,Lie1974b,Perez2007,Colle1979,Moscardo1991,%
Kraka1992,Malcolm1996,Miehlich1997,Malcolm1997,%
Leininger1997,Pollet2002,Wu1999,%
Grimme1999,Grafenstein2000,Grafenstein2005,Gutle2007,Gutle1999,%
Stoll2003,Takeda2002,Takeda2004,Gusarov2004,Gusarov2004b,%
Tsuchimochi2010,Sharkas2012,Garza2013,Garza2014,LiManni2014,%
Stoyanova2013,Fromager2015,Goodpaster2014,Hedegard2015,%
Cornaton2014,Garza2015,Garza2015b}
 in which multireference (MR) methods (\textit{e.g.}, 
complete active space or CAS)
are mixed with density functional approximations (DFAs), 
in order to  account for static and dynamic 
correlation simultaneously. 
The motivation for MR+DFT is that the MR \textit{ans\"{a}tze}
typically used in quantum chemistry
capture mostly (albeit not exclusively)
static correlation, whereas common DFAs provide 
an efficient alternative to evaluate the dynamic correlation.
Under this premise, one could think of approximating the exact 
energy as
\begin{equation}
  E \approx E^\text{MR} +  E_c^\text{DFA} [n],
  \label{eq:eq1}
\end{equation}
where $E^\text{MR}$ is the MR energy and 
$E_c^\text{DFA}$ is a (dynamic) correlation functional 
of the density $n$. 
Similarly, singlet-paired coupled cluster methods (CC0) can describe 
static correlation~\cite{Bulik2015} and hence
the incentive for CC0+DFT is analogous to that 
of MR+DFT.
Nevertheless, despite the apparent
simplicity and soundness of MR+DFT, 
these methods have not 
achieved widespread use even though concrete implementations 
have appeared in the literature since 
 \textit{circa} 1970~\cite{Lie1974a,Lie1974b}.
This is largely because of 
three problems that plague MR+DFT: 
\begin{enumerate}
  \item \textit{Double counting.}
This  problem arises because, 
in general, the MR energy contains some dynamical correlation which is 
also described by the DFA.
A raw implementation of Eq.~\ref{eq:eq1} using a
standard DFA will therefore often yield too low energies and 
unsatisfactory results~\cite{Garza2014,Garza2015b}. 
Strategies that have been proposed to avoid the double 
counting include the use of small active spaces and 
reparametrized functionals~\cite{Lie1974a,Lie1974b,Perez2007}; 
scaling the correlation
energy density by factors based on the local density 
approximation~\cite{Miehlich1997,Grafenstein2000,Grafenstein2005}
(LDA) or on the cusp condition for the pair density~\cite{Colle1979};
partition of the electron-electron interaction 
to create MR+DFT global~\cite{Sharkas2012,LiManni2014,Garza2015} 
or range-separated~\cite{Pollet2002,Stoyanova2013,Hedegard2015,%
Cornaton2014,Garza2015b} hybrids; 
separation of the correlation contributions to the 
single-particle spectrum~\cite{Gutle2007,Gutle1999};
as well partitions of the orbital 
space~\cite{Goodpaster2014,Fromager2015}
based on embedding schemes~\cite{Knizia2012,Bulik2014}.
None of these solutions is perfect: 
they either eliminate double 
counting approximately only, or eliminate it exactly but may neglect 
some static or dynamic correlation.

\item \textit{The symmetry dilemma.} 
  This refers to the fact that typical (\textit{e.g.}, Kohn--Sham (KS))
  DFAs tend to break spin symmetry in strongly correlated 
  systems~\cite{Perdew1995}  whereas MR methods do not. 
  One has therefore to choose between having physically possible
  spin densities but unphysical energies, or improved energies but
  unphysical spin densities.
  Although one may opt on retaining the symmetries of the 
  Hamiltonian, 
  this choice results in massive static correlation 
  errors for common DFAs~\cite{Cohen2008}.
  Approaches to circumvent this issue include 
  the use of functionals that 
  take as inputs not only the density but also the local 
  pair density~\cite{Moscardo1991,Miehlich1997,LiManni2014,%
    Garza2015,Garza2015b},
  or employing alternative spin densities 
  defined by a transformation 
  of the occupation numbers of the charge density 
  matrix~\cite{Perez2007,Tsuchimochi2010,Garza2013,Takeda2002,Takeda2004}. 
  The former approach is formally justified by the work of Perdew 
  \textit{et al.}~\cite{Perdew1995}, 
  but can be computationally disadvantageous
  because computing the pair density usually requires knowledge of 
  the two-particle density matrix, which can be expensive to 
  evaluate~\cite{Takeda2002,Takeda2004,Garza2015}.
  The latter approach is an \textit{ad hoc} solution to the 
  problem of computing the pair density, but suffers from 
  ambiguity~\cite{Garza2013,Garza2014} 
  in the possible definitions of the alternative densities 
  and is of course less rigorous.

\item \textit{Problems of the MR method.} 
  Typical MR techniques of quantum chemistry have 
  serious limitations. 
  High computational cost and the lack of size-consistency and 
  size-extensivity are probably the most prominent of these. 
  Of the three main problems of MR+DFT, this is the most 
  difficult to solve as it requires
  the development of novel, more efficient methods 
  for handling static correlation. 
  Alternative MR methods that have recently been 
  proposed for use in MR+DFT include 
  constrained-pairing mean-field 
  theory~\cite{Tsuchimochi2009,Tsuchimochi2010}  (CPMFT),
  projected
  Hartree--Fock~\cite{Hoyos2012,Garza2013,Garza2014} (PHF), 
  pair coupled cluster 
  doubles~\cite{Limacher2013,Limacher2014,Tecmer2014,Boguslawski2014,%
    Stein2014,Henderson2014,Garza2015,Garza2015b}
  (pCCD), and the density matrix 
  renormalization group~\cite{White1992,Hedegard2015} (DMRG). 
  While they have afforded some encouraging results, 
  none of these alternatives is flawless: 
  CPMFT+DFT often reduces (energetically) to 
  unrestricted (U)HF+DFT unless DFT exchange 
  is included in the mixture~\cite{Tsuchimochi2010}
  (which is undesirable as it introduces
  self-interaction);
  PHF lacks size consistency and extensivity~\cite{Hoyos2012};
  pCCD may not always provide a complete description 
  of the static correlation (\textit{e.g.}, 
  when dissociating N$_2$ it goes to higher energy limit 
  than UHF or CPMFT~\cite{Garza2015,Henderson2015}); 
  and DMRG still bears some 
  of the problems of traditional MR techniques 
  (\textit{e.g.}, requires definition of an active space, 
  and size-consistency can only be guaranteed by enlarging 
  this space). 
  We should also mention here that attempts have also been 
  made to add dynamic correlation to these techniques using
  approximations other than DFAs (see, \textit{e.g.}, 
  Refs.~\cite{Henderson2014,Hoyos2013,Tsuchimochi2014,%
    Limacher2014b,Jeszenszki2014,%
  Pernal2014,Pernal2015,Boguslawski2015,Limacher2015}), 
  but, save a few exceptions~\cite{Pernal2014,Pernal2015},
  the increase in cost is non-negligible in these approaches
  (and the perturbative ones can also introduce 
  instabilities~\cite{Tecmer2014,Limacher2015}).
\end{enumerate}

Here, we propose a different way to elude these
problems of MR+DFT via CC0+DFT combinations. 
The double counting is avoided by using the fact that CCD0 and BD0
have only correlations involving pairs of electrons with opposite spin; 
we show that 
the missing contributions can be added in terms of a multiple of the 
parallel-spin correlation from a density functional. 
The fact that we compute solely equal-spin correlation with the DFA
allows us then to avoid the typical problems caused by the symmetry 
dilemma without resorting to symmetry breaking. 
Furthermore, by using an appropriate meta-GGA (generalized gradient 
approximation) for the DFA,
the exactness of BD0 for two-electron systems is maintained
in BD0+DFT.
Lastly, CCD0 and BD0 are size-consistent and size-extensive, and 
have polynomial cost, rather than the combinatorial cost
of typical multiconfiguration techniques. 
Extensions incorporating RPA correlation are also derived 
and studied.

\section{Theory and Methods}

\subsection{Singlet-Paired Coupled Cluster: CCD0 and BD0}

Like other CC methods, singlet-paired coupled cluster 
doubles (CCD0) starts from an exponential \textit{ansatz}~\cite{Bulik2015}
\begin{equation}
  | \Psi_\text{CCD0} \rangle = e^{T_2^{[0]}} | \Phi_\text{RHF} \rangle,
  \label{eq:ansatz}
\end{equation}
where $| \Phi_\text{RHF} \rangle$ is a
restricted Hartree--Fock reference determinant and the
$T_2^{[0]}$ operator contains only the singlet-paired
component of the double excitations; \textit{i.e.}, if we 
let $i$ and $a$ be indices for occupied and virtual orbitals, 
respectively, $T_2^{[0]}$ is 
\begin{equation}
  T_2^{[0]} = \frac{1}{2} \sum \limits_{ijab}
  \sigma_{ij}^{ab} P_{ab}^\dag P_{ij}
  \label{eq:T2}
\end{equation}
with 
\begin{align}
  P_{ij}  & = \frac{1}{\sqrt{2}}  \left( c_{j\uparrow} 
  c_{i\downarrow} +   c_{i\uparrow}  c_{j\downarrow} \right) 
    \label{eq:Pab} \\
  & = \frac{1}{\sqrt{2}}  \left( c_{j\uparrow} 
  c_{i\downarrow} - c_{j\downarrow} c_{i\uparrow}   \right). \nonumber 
\end{align}
To shed light on the idea behind CCD0 (and this is 
helpful for understanding the justification for our CCD0+DFT 
combinations), 
it is instructive 
to consider standard CCD. 
The full double excitations cluster operator 
$T_2$ used in the latter can be expressed 
as~\cite{Geertsen1986,Scuseria1988,Piecuch1990}
\begin{equation}
  T_2 = T_2^{[0]} + T_2^{[1]}, 
\end{equation}
where $T_2^{[0]}$ is defined above and 
\begin{equation}
  T_2^{[1]} = \frac{1}{2} \sum \limits_{ijab}
  \pi_{ij}^{ab} \mathbf{Q_{ab}^\dag} \cdot 
  \mathbf{Q_{ij}}
  \label{eq:T21}
\end{equation}
where $\mathbf{Q_{ij}}$ may be written as a vector 
$\mathbf{Q_{ij}} = ( Q_{ij}^{+}, Q_{ij}^{0}, Q_{ij}^{-} )^t$
whose components are
\begin{equation}
  Q_{ij}^{+} =   c_{j\uparrow}  c_{i\uparrow} , \quad 
  Q_{ij}^{-} =   c_{j\downarrow}  c_{i\downarrow} ,
  \label{eq:Q1}
\end{equation}
\begin{align}
  Q_{ij}^0 & = \frac{1}{\sqrt{2}}  \left( c_{j\uparrow} 
  c_{i\downarrow} -   c_{i\uparrow}  c_{j\downarrow} \right) 
  \label{eq:Q0} \\
  & = \frac{1}{\sqrt{2}}  \left( c_{j\uparrow} 
  c_{i\downarrow} +   c_{j\downarrow} c_{i\uparrow} \right). \nonumber
\end{align}
Note that $P_{ij}^\dag$ acting on an empty state
$ | \, \rangle$ gives a singlet wavefunction
(hence the name of singlet-paired coupled cluster), whereas
$\mathbf{Q_{ij}^{\dag}} | \, \rangle$ yields pure triplet 
components with different $m$.
We see then that CCD incorporates contributions from both the singlet- and
triplet-paired components of $T_2$.
Thus, more correlation is taken into account in CCD as compared to CCD0. 
However, it is the combination of the singlet- and
triplet-paired components what causes the failure
of CCD (and other common CC approximations) in strongly correlated 
systems, which results in the correlation becoming too large
or even complex~\cite{Bulik2015}.
By removing $T_2^{[1]}$ from
$T_2$,
CCD0 relinquishes a fraction of the correlation 
in exchange for safeguard against this breakdown.
One may also choose to retain only the triplet-paired 
component in order to avoid the breakdown, but this alternative
recovers less correlation than CCD0~\cite{Bulik2015}.

Just as CCD can be improved by using (approximate) 
Brueckner orbitals in the Brueckner doubles 
method~\cite{Dykstra1997,Handy1989,Kobayashi1991} (BD), 
CCD0 has an analogous extension in singlet-paired 
BD (BD0). Recall that the exact Brueckner orbitals are those that 
make the coefficients of singly-substituted determinants in the full
configuration interaction (FCI) wavefunction vanish~\cite{Sherrill1998}.
The BD0 \textit{ansatz} is thus very similar to that of 
Eq.~\ref{eq:ansatz}
\begin{equation}
  | \Psi_\text{BD0} \rangle = e^{T_2^{[0]}} | \Phi_\text{BD} \rangle,
  \label{eq:bd0}
\end{equation}
except that the RHF reference is replaced by 
a determinant with approximate Brueckner orbitals $| \Phi_\text{BD} \rangle$
which zero out the amplitudes of single substitutions in a 
subspace of single and double substitutions. 
We remark that a singlet-paired CCSD method (CCSD0) is possible 
too~\cite{Bulik2015}, 
but (just as it happens for BD and CCSD~\cite{Scuseria1987})
results from CCSD0  are not significantly different from those of BD0.

The CCD0 and BD0 methods have certain advantages over traditional MR 
techniques. Both algorithms scale as $\mathcal{O}(N^6)$; 
the cost of CCD0  is identical to that of CCD
and the same holds for BD0 and BD.
While not precisely inexpensive, it is certainly better than 
the combinatorial scaling of typical MR techniques and, as 
CC-based methods, CCD0 and BD0 may be able to take advantage of recently
developed approaches~\cite{Liakos2015}
to decrease the cost of coupled cluster calculations. 
In addition, the use of exponential wavefunctions (Eqs.~\ref{eq:ansatz} and 
\ref{eq:bd0}) guarantees size extensivity. 
Size consistency is also satisfied by the 
exponential \textit{ansatz} provided that the reference 
determinant be size consistent. 
Lastly, it is easy to modify existing CCD and BD 
subroutines to implement CCD0 and BD0, respectively. 
Basically, most restricted
CCD codes use as fundamental variables 
the $t^{a\uparrow b\downarrow}_{i\uparrow j\downarrow}$ 
amplitudes of the full double excitations cluster operator 
$T_2$, which may be expressed as~\cite{Scuseria1988}
\begin{equation}
  T_2 = \frac{1}{2} \sum \limits_{ijab} 
  t^{a\uparrow b\downarrow}_{i\uparrow j\downarrow}
  \sum \limits_{\alpha \beta} c^\dag_{a\alpha} 
  c^\dag_{b\beta} c_{j\beta} c_{i\alpha} , 
  \label{eq:T2b} 
\end{equation}
and, because the singlet-paired
(triplet-paired)
component of $T_2$ is 
symmetric (antisymmetric)
with respect to the interchange of $i$ and $j$, 
the $\sigma_{ij}^{ab}$ amplitudes (Eq.~\ref{eq:T2})
of the singlet-paired component obey the following relations
\begin{equation}
  \sigma_{ij}^{ab}  = \sigma_{ij}^{ba} = 
  \sigma_{ji}^{ab} = \sigma_{ji}^{ba}  = 
  \frac{t^{a\uparrow b\downarrow}_{i\uparrow j\downarrow} + 
    t^{b\uparrow a\downarrow}_{i\uparrow j\downarrow}}{2},
  \label{eq:sigmas}
\end{equation}
which implies that we can implement CCD0 by replacing 
$t^{a\uparrow b\downarrow}_{i\uparrow j\downarrow}$ by 
$1/2(t^{a\uparrow b\downarrow}_{i\uparrow j\downarrow} + 
t^{b\uparrow a\downarrow}_{i\uparrow j\downarrow})$ at each 
iteration in 
a CCD code. Analogous remarks apply for 
the implementation of BD0 starting from BD subroutines.

The above discussion regarding CCD0 and BD0
suffices for the purposes of this paper. 
The interested reader may found
further details about singlet-paired coupled cluster 
in Ref.~\cite{Bulik2015}.

\subsection{Adding DFT Correlation to CC0
Without Double Counting: CC0+DFT}

To combine CCD0 and BD0 with density functionals while avoiding 
double counting, we first note that the CCD0 wavefunction can 
be expanded as
\begin{align}
  e^{T_2^{[0]}} | \Phi_\text{RHF} \rangle & = 
  | \Phi_\text{RHF} \rangle + T_2^{[0]} | \Phi_\text{RHF} \rangle +
  \frac{T_2^{[0]} T_2^{[0]}}{2!} | \Phi_\text{RHF} \rangle 
  + \cdots 
    \nonumber \\
  & =  | \Phi_\text{RHF} \rangle + 
  \sum \limits_{ijab} \sigma_{ij}^{ab}  
  c_{a\uparrow}^\dag  c_{b\downarrow}^\dag
  c_{j\downarrow} c_{i\uparrow}  | \Phi_\text{RHF} \rangle +  \cdots, 
   \label{eq:expand}
\end{align}
and that the CCD0 correlation energy is
\begin{equation}
  E_c^\text{CCD0} = \sum \limits_{ijab}
  \sigma_{ij}^{ab} v^{a\uparrow b\downarrow}_{i\uparrow j\downarrow}, 
  \label{eq:Ec1}
\end{equation}
where $v^{a b}_{i j} = \langle i j |  a b \rangle$
is a two-electron integral in the Dirac notation.
It is clear from these expressions that CCD0 neglects 
contributions to the correlation from 
equal-spin excitations: only integrals involving pairs of 
opposite-spin electrons 
in occupied or virtual orbitals are 
weighted in the correlation energy.
The same applies to BD0 except, of course, that the orbitals 
$i$, $j$, $a$, and $b$ would be (approximate) Brueckner, rather than RHF.
One can thus think of adding equal spin correlation to 
CCD0 using a density functional; 
the correlation of the parallel-spin density given by a functional 
contains, in principle, contributions from all excitations 
involving same-spin electrons only. 
Hence, we may, in the spirit of Eq.~\ref{eq:eq1}, write
\begin{equation}
  E_c^\text{CCD0+pDFT} = E_c^\text{CCD0} + 
  E_{c \, \uparrow \uparrow}^\text{DFA} [n_\uparrow, n_\downarrow] + 
  E_{c \, \downarrow \downarrow}^\text{DFA} [n_\uparrow, n_\downarrow],
  \label{eq:Ec2}
\end{equation}
where the ``p'' in pDFT is for ``\textit{parallel-spin}'' and 
$E_{c \, \alpha \alpha}^\text{DFA}$ is the correlation of the 
spin-$\alpha$ density computed by a DFA. 
To evaluate $E_{c \,\alpha \alpha}^\text{DFA}$, we can use 
the exchange-like spin resolution of the correlation 
by Stoll \textit{et al.}~\cite{Stoll1978}
\begin{equation}
  E_{c \,\uparrow \uparrow}^\text{DFA} [n_\uparrow, n_\downarrow] + 
  E_{c \,\downarrow \downarrow}^\text{DFA} [n_\uparrow, n_\downarrow]  =
  E_c^\text{DFA} [n_\uparrow, 0] +  E_c^\text{DFA} [0,n_\downarrow]. 
  \label{eq:Stoll}
\end{equation} 
Also, because CCD0 and BD0 do not break spin symmetry and 
we will be working only with closed shells, the total 
CCD0+pDFT energy simplifies to
\begin{equation}
  E^\text{CCD0+pDFT}_\text{total} = E^\text{CCD0} + 
  2  E_c^\text{DFA} [n_\uparrow, 0], 
  \label{eq:Etot}
\end{equation} 
and likewise for BD0.

Now, looking back at the triplet-paired component of $T_2$ that 
is excluded in CCD0 (Eqs.~\ref{eq:T21}, \ref{eq:Q1}, and 
\ref{eq:Q0}), we realize that,  apart from the same-spin correlation,
CCD0 misses opposite-spin contributions from the $m = 0$ component of 
$T_2^{[1]}$ ($Q_{ij}^0$ in Eq.~\ref{eq:Q0}). 
However, for a closed shell, the 
$m = +1$, $m = 0$, and $m = -1$ channels of $T_2^{[1]}$
contribute all equally to the energy. 
In other words, 
the spin-up correlation associated with the $m = +1$ element
of $T_2^{[1]}$ is, by symmetry, identical to the opposite-spin
correlation of the $m =0$ part. 
We may therefore incorporate the opposite-spin 
energy that is missing
in CCD0+pDFT by adding $E_c^\text{DFA} [n_\uparrow, 0]$ once more to
the total energy 
\begin{equation}
  E^\text{CCD0+tDFT}_\text{total} = E^\text{CCD0} + 
  3  E_c^\text{DFA} [n_\uparrow, 0], 
  \label{eq:Etot2}
\end{equation} 
where the ``t'' in tDFT indicates that the full contributions from
the \textit{triplet-paired} component of $T_2$ are being 
taken into account. 
Again, the above analysis applies to BD0 too.

In this work, we
employ Eqs.~\ref{eq:Etot} and \ref{eq:Etot2} in a 
non-self-consistent manner:
after a self-consistent
CCD0 (BD0) calculation, the equal spin 
correlation is obtained in a single-shot evaluation of a DFA
using the densities from the reference RHF (Brueckner) determinant.
This is a reasonable assumption because, when adding DFA dynamic 
correlation to MR wavefunctions, the error in the approximate 
functional is often larger than the error due to lack of
self-consistency~\cite{Stoll1985}. 
We benchmark both CC0+pDFT and CC0+tDFT with CC0 referring 
to both CCD0 and BD0. 
The motivation for testing both CC0+pDFT and CC0+tDFT is that 
if the $\sigma_{ij}^{ab}$ amplitudes are close to 
those of the full configuration-interaction (FCI), and the 
same-spin DFA correlation is fairly accurate,  then
we would anticipate CC0+tDFT to outperform 
CC0+pDFT. However, if it is not the case that the $\sigma_{ij}^{ab}$ 
amplitudes of CC0
are similar to the exact ones, then 
CC0+tDFT may overcorrelate and not necessarily be better than 
CC0+pDFT.

\subsection{Eluding the Symmetry Dilemma and Selecting
an Adequate Density Functional}

So far we have identified the advantages of CC0 over traditional 
MR techniques, and 
the derivation of CC0+DFT above explains how double counting 
is avoided; we are just left to deal with the symmetry dilemma.
To illustrate how this issue can be avoided, consider the 
paradigm for static correlation in quantum chemistry: 
the H$_2$ singlet molecule at the dissociation 
limit~\cite{Cohen2008}.
For the exact, symmetry-adapted, wavefunction, 
$n_\uparrow$ and $n_\downarrow$ 
are distributed equally in each atom, but the probability 
of finding both electrons at the same atom is zero. 
That is, the correlation is purely static and there is no
dynamic correlation because the electrons are never close to each other. 
However,
with symmetry-adapted densities, common DFAs yield a
nonzero $E_c^\text{DFA}$ that is caused (largely or completely) by the 
fact that semilocal approximations depending on the density and 
its gradients ``see'' the densities of two different
electrons on top of each other; they do not know that the local 
pair density is zero everywhere. 
This problem is normally alleviated by breaking spin symmetry in a 
way that localizes an electron of a certain spin around a 
specific nucleus---so that 
the functional won't see the densities of different electrons 
overlapping---or by introducing functionals that depend not only 
on the density, but also on the local pair density~\cite{Perdew1995}. 

Because we evaluate the residual 
correlation in terms of parallel-spin correlation only
(see Eqs.~\ref{eq:Etot} and \ref{eq:Etot2}), 
the spurious opposite-spin correlation due to the use 
of symmetry-adapted densities when describing 
the dissociation limit of H$_2$ (and
stretched bonds in general)
disappears in CC0+DFT.  
Therefore, there is no need for symmetry breaking or using the
local pair density.
Nonetheless, we should still consider the problem of self-interaction
in approximate functionals,
which produces artificial self-correlation and is present in
the local density approximation (LDA) and generalized gradient 
approximations (GGAs). 
One can get rid of this issue, at least in the case of two-electron 
systems, by using meta-GGA functionals: DFAs that depend 
on the density, its gradient, and the kinetic energy density. 
We have therefore chosen to test here combinations of CC0 with 
two meta-GGAs: the Tao--Perdew--Staroverov--Scuseria~\cite{Tao2003}
(TPSS) functional, 
and the recently developed 
strongly constrained and appropriately normed (SCAN) 
functional of Sun \textit{et al.}~\cite{Sun2015}.
These functionals were selected based on the facts that 
(1) they are nonempirical; (2) they are free of one-electron 
self-interaction error, so that they yield no same-spin 
correlation for two-electron singlets and both BD0+TPSS and 
BD0+SCAN will be exact for these systems; and (3) that despite 
these similarities, TPSS and SCAN are quite different because 
they were designed based on different paradigms. 
Specifically, TPSS respects the paradigms of describing correctly
one- or two-electron densities and slowly varying 
densities~\cite{Tao2003}, 
whereas SCAN obeys all the 17 known exact constrains that a 
meta-GGA can obey, and is exact or near exact for a set 
of rare-gas atoms and certain nonbonded interactions~\cite{Sun2015}
(the so-called ``appropriate norms'').

\subsection{Improving the Spin Resolution of the SCAN correlation}

In Section 2.2 we used the 
spin resolution of $E_c^\text{DFA}$
by Stoll \textit{et al.}~\cite{Stoll1978}
($E_{c \, \uparrow \uparrow}^\text{DFA} [n_\uparrow, n_\downarrow] = 
E_c^\text{DFA} [n_\uparrow, 0]$)
to extract the parallel-spin correlation. 
However, this decomposition is correct only 
for fully spin-polarized densities and 
in the high-density limit of the uniform electron gas~\cite{Gori-Giorgi2004}.
In other regimes of the uniform gas, it
 exaggerates the equal-spin correlation.
It is possible, however, because of how the correlation is 
composed in SCAN, 
to formulate a better educated guess for the spin resolution 
of this functional. 

The dependence on the kinetic energy density, 
$\tau(r) = \sum_{i}^\text{occ} (1/2) | \nabla \ \psi_i(r) |^2$, is 
introduced in SCAN via the dimensionless variable $\alpha$~\cite{Sun2015}:
\begin{equation}
  \alpha = (\tau - \tau^W)/\tau^\text{unif} > 0, 
\end{equation}
where $\tau^W = | \nabla n | ^2/8n$ is the single-orbital limit 
of $\tau$, and $\tau^\text{unif}$ is the uniform density limit. 
SCAN constructs the correlation energy density $\varepsilon_c$
by interpolation and extrapolation from the 
$\alpha = 0$ ($\varepsilon_c^0$) and $\alpha = 1$
($\varepsilon_c^1$) limits. 
It is therefore natural to build its spin resolution in the 
same way:
\begin{equation}
  \varepsilon_c^{\uparrow \uparrow} = 
  \varepsilon_c^{1 \, \uparrow \uparrow}  + 
  f_c(\alpha) \left[ \varepsilon_c^{0 \, \uparrow \uparrow}  - 
    \varepsilon_c^{1 \, \uparrow \uparrow}  \right], 
\end{equation}
where $f_c(\alpha)$ 
(which is given in the Supporting Information of Ref.~\cite{Sun2015}) 
satisfies $f_c(0) = 1$, $f_c(1) = 0$, and $f_c(\infty) = -0.70$, 
and likewise for the spin-down component.
We note that $\varepsilon_c^{0 \, \uparrow \uparrow} = 
\varepsilon_c^{0 \, \downarrow \downarrow} = 0$, since there is no
parallel-spin correlation for two electrons in the 
same spatial orbital.
Thus, the fraction of same-spin correlation in SCAN
depends only on the fraction of same-spin correlation in the uniform 
density limit, $\varepsilon_c^{1 \, \uparrow \uparrow} +
\varepsilon_c^{1 \, \downarrow \downarrow} = 
(1-F_{\uparrow \downarrow}) \varepsilon_c^1$.
The total parallel-spin correlation energy density thus becomes
\begin{equation}
    \varepsilon_c^{\uparrow \uparrow} + 
  \varepsilon_c^{\downarrow \downarrow} = 
  \left( 1 - f_c(\alpha) \right) \left( 1-F_{\uparrow \downarrow} \right)
  \varepsilon_c^1 .
  \label{eq:eq20}
\end{equation}
All of the terms in the above equation are already determined in 
the original 
SCAN functional except for $F_{\uparrow \downarrow}$, 
the fraction of opposite-spin correlation density in the uniform 
density limit.
Gori-Giorgi and Perdew~\cite{Gori-Giorgi2004}
have worked out the spin resolution for the 
uniform electron gas: they determined fractions 
$F_{\sigma \sigma'}$ such that 
$\varepsilon_c^{\sigma \sigma'} = \varepsilon_c F_{\sigma \sigma'}$
in terms of the local Wigner--Seitz radius 
$r_s = (4 \pi n/3)^{-1/3}$ and the spin polarization 
$\zeta = (n_\uparrow - n_\downarrow)/n$.
Therefore, we calculate 
$F_{\uparrow \downarrow} (r_s , \zeta )$ using Equation 9 of 
Ref.~\cite{Gori-Giorgi2004}, which is an interpolation between 
exact results for the high ($r_s \rightarrow 0$) and low 
($r_s \rightarrow \infty$) density limits that agrees with 
available quantum Monte Carlo data~\cite{Gori-Giorgi2000}
in the range $0.8 \leq r_s \leq 10$ with $\zeta = 0$.

Note that integrating Eq.~\ref{eq:eq20} yields all the SCAN
correlation that needs to be added in CC0+pDFT; 
for CC0+tDFT, this energy should be multiplied by 
3/2 (see Section 2.2).
Also, Eq.~\ref{eq:eq20} gives no correlation for 
two-electron singlets because, for these systems,
$\alpha = 0$ and $f_c(0) = 1$. 
Thus, BD0+SCAN remains exact for two-electron systems 
when the parallel-spin correlation is resolved
using Eq.~\ref{eq:eq20}.

\subsection{Modeling the Long-Range Correlation with the 
Random Phase Approximation}

Adding residual correlation to CC0 via semilocal DFAs
has the following pitfall: 
Semilocal functionals can't capture the long-range part 
of the correlation in van der Waals interactions~\cite{Tao2012,Klimes2012}. 
Therefore, the description of these forces in
CC0+DFT is essentially the same as that of CC0. 
Because van der Waals forces are very sensitive to dynamic 
correlation, and CC0 misses a substantial part of it
(the triplet pairing channel, to be precise), 
CC0+DFT may be inadequate for simulating this sort of
interactions. 

This problem can be solved by the technique of 
range-separation~\cite{Savin1995,Ikura2001,Yanai2004,Vydrov2006a,%
Vydrov2006b,Janesko2009a}: 
The interelectron Coulomb operator $r_{12}^{-1}$ is separated
into a short-range (SR) component and its long-range (LR) complement
\begin{equation}
\frac{1}{r_{12}}=\underbrace{\frac{1 - \mathrm{erf}(\mu r_{12})}{r_{12}}}%
_\text{SR}+\underbrace{\frac{\mathrm{erf}(\mu r_{12})}{r_{12}}}_\text{LR} ,
\label{eq:split}
\end{equation}
where erf is the error function and $\mu$ a
parameter defining the range separation. (Other partitions 
of $r_{12}^{-1}$ are possible, but the above choice is most 
convenient as it has been used to parametrize 
screened functionals and also facilitates 
the evaluation of two-electron integrals.)
We can then evaluate the DFA correlation with the SR interaction only;
the LR complement should 
be computed with an approximation capable of describing the long-range 
part of the van der Waals forces in a way that does not 
add double counting to CC0 (and, desirably, that won't increase 
the cost). 
As we shall explain shortly, 
a suitable approach that satisfies all of these 
requirements consists in applying the direct
random phase approximation (dRPA) in a manner 
similar to that by Janesko \textit{et al.}~\cite{Janesko2009}.

For the sake of clarity, let us consider 
how the RPA models correlation.
The RPA requires the solution of an eigenvalue equation of the 
form~\cite{Scuseria2008,Janesko2009,Eshuis2012}
\begin{equation}
\begin{pmatrix}
\mathbf{A} & \mathbf{B} \\ 
 -\mathbf{B} & \mathbf{-A}
\end{pmatrix}
\begin{pmatrix}
\mathbf{X} & \mathbf{Y} \\ 
\mathbf{Y} & \mathbf{X}
\end{pmatrix}
=
\begin{pmatrix}
\mathbf{X} & \mathbf{Y} \\ 
\mathbf{Y} & \mathbf{X}
\end{pmatrix}
\begin{pmatrix}
\boldsymbol{\omega} & 0 \\
0 & -\boldsymbol{\omega} 
\end{pmatrix}, 
\label{eq:matrix}
\end{equation}
where $\mathbf{A}$, $\mathbf{B}$, $\mathbf{X}$, 
and $\mathbf{Y}$ are all of dimension $ov \times ov$, with $o$ and 
$v$ being the number of occupied and virtual spinorbitals, 
respectively. 
In the full RPA, 
the elements of $\mathbf{A}$ and $\mathbf{B}$  are
\begin{equation}
  A_{ia,jb} = ( \epsilon_a - \epsilon_i ) \delta_{ij} \delta_{ab}
  + \langle ib || aj \rangle ,
\label{eq:A}
\end{equation}
\begin{equation}
  B_{ia,jb} = 
   \langle ij || ab \rangle.
\label{eq:B}
\end{equation}
Equations \ref{eq:matrix}--\ref{eq:B} are used 
for the calculation of excited states. 
To obtain the ground state correlation energy, one notes that 
the Tamm--Dancoff approximation (also known as configuration interaction 
singles) also computes 
excited states with Eq.~\ref{eq:matrix}
but setting $\mathbf{B} = 0$
(\textit{i.e.}, it solves 
for $\mathbf{A} \mathbf{Z} = \mathbf{Z} \boldsymbol{\nu}$). 
While this approximation contains only excitation operators, 
the RPA also includes de-excitation operators 
that can be thought of as correlating the ground state.
Thus, 
the RPA correlation energy for the ground state is normally 
written as~\cite{Scuseria2008,Furche2008}
\begin{equation}
  E_c^\text{RPA} = \frac{1}{2} \text{Tr}(\boldsymbol{\omega} - 
  \mathbf{A}).
  \label{eq:EcRPA}
\end{equation}
In the dRPA, the correlation is given by the 
same expression, except the the exchange contributions are 
neglected~\cite{Scuseria2008,Janesko2009,Eshuis2012}
 (\textit{i.e.}, $\langle pq || rs \rangle$ is
replaced by  $\langle pq | rs \rangle$ 
in Eqs.~\ref{eq:A} and \ref{eq:B}). 
The dRPA has been most useful for incorporating correlation 
in methods that already contain exchange (see, \textit{e.g.}, 
Refs.~\cite{Scuseria2008,Janesko2009,Toulouse2011} and references
therein). 
Direct RPA also has the enormous advantage that the correlation is 
guaranteed to be real if the orbitals obey
 the \textit{aufbau} principle, 
which is not true for the full RPA~\cite{Seeger1977}.
As a matter of fact, the full RPA correlation 
becomes complex in the presence of an RHF 
instability~\cite{Furche2001,Cui2013} (negative eigenvalue in the Hessian).
The dRPA does not suffer any of the instability problems discussed
above for full RPA or CCD because of the neglect of exchange.
This minimalist description of the RPA correlation is sufficient 
for our purposes here; the interested reader is referred 
to reviews on the subject~\cite{Eshuis2012,Furche2008} for 
further details.

\renewcommand{\arraystretch}{1.35}
\begin{table*}
  \begin{center}
    \caption{Summary of the CC0+DFT methods tested here. 
      The notation, closed-shell energy formulas, and 
      relevant equations are given;
      CC0 can refer to CCD0 or BD0 and the densities are 
      from the RHF or Brueckner reference determinants, respectively; 
      the ``p'' in pDFT is for \textit{parallel spin}; the ``t'' in tDFT 
      is for \textit{triplet-pairing component};
      $E_{c \, \uparrow \uparrow}^\text{sr-rSCAN}$ is the spin-up 
      short-range SCAN correlation using the spin
      \textit{resolution} of Section 2.4;
      $E_{c \, \uparrow \uparrow}^\text{lr-dRPA}$ 
      is the spin-up long-range dRPA correlation;
      and the rest of the notation is given in the text.}
    \label{tab:methods}
    \scalebox{0.82}{
      \begin{tabular*}{1.1\textwidth}{@{\extracolsep{\fill}}  lll  }
        \hline
        Method & Energy Formula  & Relevant Eqs. \\
        \hline 
        CC0+$E_c^\text{DFT}$ & $ E^\text{CC0} + 
        E_c^\text{DFA} [n_\uparrow, n_\downarrow ]$ & See Sec. 2.6. \\
        CC0+pDFT & $ E^\text{CC0} + 
        2 E_c^\text{DFA}[n_\uparrow,0]$ & \ref{eq:Ec2}, \ref{eq:Stoll} \\
        CC0+tDFT & $ E^\text{CC0} + 
        3 E_c^\text{DFA}[n_\uparrow,0]$ & \ref{eq:Stoll}, \ref{eq:Etot2} \\
        CC0+prSCAN & $ E^\text{CC0} + 
        2 E_{c \, \uparrow \uparrow}^\text{rSCAN} [n_\uparrow, n_\downarrow ]  $
        & \ref{eq:Ec2}, \ref{eq:eq20} \\
        CC0+trSCAN & $ E^\text{CC0} + 
        3 E_{c \, \uparrow \uparrow}^\text{rSCAN} [n_\uparrow, n_\downarrow ]  $
        & \ref{eq:Etot2}, \ref{eq:eq20}  \\
        LC-CC0+pDFT & $ E^\text{CC0} + 
        2 E_c^\text{sr-DFA} [n_\uparrow, 0]
        + 2 E_{c \, \uparrow \uparrow}^\text{lr-dRPA}$ & 
         \ref{eq:Ec2}, \ref{eq:Stoll}, 25--31  \\
        LC-CC0+tDFT & $ E^\text{CC0} + 
        3 E_c^\text{sr-DFA} [n_\uparrow, 0]
        + 3 E_{c \, \uparrow \uparrow}^\text{lr-dRPA}$ & 
         \ref{eq:Etot2}, \ref{eq:Stoll}, 25--31 \\
        LC-CC0+prSCAN & $ E^\text{CC0} + 
        2 E_{c \, \uparrow \uparrow}^\text{sr-rSCAN} [n_\uparrow, n_\downarrow ] 
        + 2 E_{c \, \uparrow \uparrow}^\text{lr-dRPA}$ & 
         \ref{eq:Ec2}, \ref{eq:eq20}, 25--31 \\
        LC-CC0+trSCAN & $ E^\text{CC0} + 
        3 E_{c \, \uparrow \uparrow}^\text{sr-rSCAN} [n_\uparrow, n_\downarrow ] 
        + 3 E_{c \, \uparrow \uparrow}^\text{lr-dRPA}$ & 
         \ref{eq:Etot2}, \ref{eq:eq20}, 25--31 \\
 \hline    
      \end{tabular*}
    }
  \end{center}
\end{table*}
\renewcommand{\arraystretch}{1.0}

From this explanation, and the analysis in Section 2.2,
it is straightforward to see how 
long-range dRPA correlation may be added 
to CC0 without double counting. 
The correlation is still expressed by Eq.~\ref{eq:EcRPA}; 
only the elements of $\mathbf{A}$ and $\mathbf{B}$ need to 
be altered:
\begin{equation}
  A_{ia,jb}
  = ( \epsilon_a - \epsilon_i ) \delta_{ij} \delta_{ab}
  +  \langle ib | v_{ee}^\text{lr}| aj \rangle ,
\label{eq:Anew}
\end{equation}
\begin{equation}
  B_{ia,jb} = 
\delta_{\sigma_i \sigma_j} 
  \langle ij | v_{ee}^\text{lr}| ab \rangle ,
\label{eq:Bnew}
\end{equation}
where $\delta_{\sigma_i \sigma_j} = 1$ if the spin functions 
of the spinorbitals $\chi_i(\mathbf{x})$ and $\chi_j(\mathbf{x})$
are identical and $\delta_{\sigma_i \sigma_j} = 0$ otherwise, 
and $ \langle ij | v_{ee}^\text{lr}| ab \rangle $ indicates 
that the two-electron integral be evaluated with the long-range
interaction
\begin{align}
  \langle ij | v_{ee}^\text{lr}| ab \rangle    = 
  \int & d\mathbf{x}_1 d\mathbf{x}_2
  \chi_i^* (\mathbf{x}_1) \chi_j^* (\mathbf{x}_2)
  \frac{\text{erf}(\mu r_{12})}{r_{12}}  \times \nonumber \\
  &  \chi_a(\mathbf{x}_1) \chi_b(\mathbf{x}_2). 
\label{eq:IntNew}
\end{align}
By calculating the dRPA correlation with the $\mathbf{A}$
and $\mathbf{B}$ matrices as defined in Eqs.~\ref{eq:Anew} and 
\ref{eq:Bnew},
only parallel spin correlation that does not overlap with 
the CC0 correlation is obtained. 
This is the same strategy that we utilized in Section 2.2
to avoid the double 
counting between DFT and CC0.
Therefore, the same considerations discussed in Section 2.2 
apply; one can add to CC0 only the parallel-spin correlation,
or the full triplet-paired component contributions by multiplying 
the total equal-spin correlation by 3/2.

The RPA correlation describes correctly dispersion and van der Waals 
interactions~\cite{Angyan2005,Dobson2006}, 
and is exact for long-range correlations~\cite{Yan2000}. 
In addition,
Scuseria \textit{et al.}~\cite{Scuseria2008}
have shown that solving the RPA eigenvalue 
problem of Eq.~\ref{eq:matrix} is equivalent to solving for 
$\mathbf{T} = \mathbf{Y} \mathbf{X}^{-1}$ in a Riccati
CCD equation 
\begin{equation}
  \mathbf{B} + \mathbf{AT} + \mathbf{TA} + 
  \mathbf{TBT} = 0,
  \label{eq:rpa2}
\end{equation}
and that $E_c^\text{RPA}$ in Eq.~\ref{eq:EcRPA} can also be
expressed as
\begin{equation}
  E_c^\text{RPA} = \frac{1}{2} \text{Tr}(
  \mathbf{BT}).
  \label{eq:EcRPA2}
\end{equation}
Using the Cholesky decomposition of $\mathbf{A}$ and $\mathbf{B}$,
Equation \ref{eq:rpa2} can be solved in $\mathcal{O}(N^4)$ 
computational effort~\cite{Scuseria2008,Janesko2009}, 
where $N$ is the number of basis functions. 
This does not exceed the cost of CCD0. 
Hence, the dRPA correlation fulfills all of the requirements that we 
were looking for: it can be added to CCD0 without double counting 
or increase in scaling, and it describes properly long-range 
interactions.
As we do for $E_c^\text{DFA}$, 
the dRPA correlation is obtained from  a 
single-shot, post-SCF, calculation using the RHF or BD orbitals 
and added to the CC0 energy.

As mentioned above, $E_c^\text{DFA}$ needs to be evaluated 
with the SR interaction only, in order to avoid 
double counting with the dRPA correlation.
To the best of our knowledge, 
there are no parametrizations for the short-range
correlation of TPSS or SCAN. However, codes for the
range-separated parametrization of the LDA are 
available~\cite{Paziani2006}. 
Therefore, we use an approximate local scaling to estimate
the SR meta-GGA same-spin correlation
\begin{align}
  E_{c \, \uparrow \uparrow}^{\text{sr},\mu}[n] = 
  \int n_{\uparrow} (r) 
  \frac{\epsilon_{c,\mu}^{\text{sr-LDA}}(n_\uparrow,0)
  }{\epsilon_c^\text{LDA}(n_\uparrow,0)} 
   \epsilon_{c \, \uparrow \uparrow}^\text{MGGA}(n,\nabla n, \tau) d^3r,
   \label{eq:eq8}
\end{align}
where $\epsilon_{c,\mu}^\text{sr-DFA}$ and
$\epsilon_c^\text{DFA}$ are 
short- and full-range DFA correlation energy densities,
respectively, taking spin densities as inputs.
A similar LDA-based scaling has been shown to produce reasonable
results in our recent work on range-separated hybrids of pCCD and 
density functionals~\cite{Garza2015b}. 
Note that we have used the decomposition by 
Stoll \textit{et al.}~\cite{Stoll1978} (Eq.~\ref{eq:Stoll})
for determining the effect of considering only parallel 
spin correlation on the scaling factor; the 
Gori-Giorgi--Perdew~\cite{Gori-Giorgi2004}
spin resolution can't be used for this because 
the $F_{\sigma \sigma'}$ fractions (see Section 2.4)
are independent of $\mu$.
Nonetheless, $\epsilon_{c \, \uparrow \uparrow}^\text{MGGA}$ 
may still be evaluated with the spin resolution 
of Eq.~\ref{eq:eq20} for the SCAN functional. 
Also, both $\epsilon_{c,\mu}^\text{sr-LDA}(n_\uparrow, 0)$ and
$\epsilon_c^\text{LDA}(n_\uparrow, 0)$ are expected to exaggerate 
equal-spin correlation, resulting in error 
cancellation on the ratio that defines the scaling.
Lastly, Eq.~\ref{eq:eq8} maintains the 
$E_{c \, \uparrow \uparrow}^{\text{sr},\mu} = 0$
condition
for two-electron singlets as long as 
$\epsilon_{c \, \uparrow \uparrow}^\text{MGGA}$ be 
evaluated with an adequate meta-GGA.

The separation of the electron-electron interaction with 
Eq.~\ref{eq:split} also requires one to define a range-separation 
parameter $\mu$. 
In the case of standard LC-KS-DFT, where Eq.~\ref{eq:split} is 
used to divide the exchange interaction only, the optimal 
$\mu$ is highly system and property
dependent~\cite{Stein2009,Abramson2011,Kronik2012,%
Autschbach2014,Garza2014a}.
However, here we are interested in splitting the correlation 
in SR and LR, and in this case $\mu$ is more universal:
Fromager \textit{et al.}~\cite{Fromager2007,Fromager2009}
have demonstrated, based on 
physical arguments and numerical experiments, that the optimal $\mu$ 
for evaluating the SR correlation with semilocal DFAs is in the vicinity 
of  $\mu \approx 0.4$ au. Therefore, we set $\mu = 0.4$ au in 
all of our calculations (note that this value is widely used 
in LC-KS-DFT functionals such as, \textit{e.g.}, 
LC-$\omega$PBE~\cite{Vydrov2006b}).
Additionally,  
by fixing $\mu$, size consistency and extensivity are 
preserved; system-dependent definitions of $\mu$ not all 
can guarantee this~\cite{Garza2014a,Karolewski2013}. 
We note that it is also possible to define a physically-motivated 
local  $\mu(r)$, but this has the caveat that 
locally-range separated hybrids are difficult and expensive to 
evaluate~\cite{Krukau2008,Henderson2009b}. 

Lastly, we remark that range separation using long-range dRPA 
correlation is one way to introduce the long-range part of the 
van der Waals interaction in CC0+DFT; other approaches 
are possible. For example, specialized van der Waals 
functionals of varying degree of empiricism have been developed
by various groups (see, \textit{e.g.}, 
Refs.~\cite{Dion2004,Lee2010,Vydrov2009,Sato2009} and 
references therein). 
These can be used in CC0+DFT as long as meaningful spin resolution 
for the correlation of the functional exist.

\subsection{Summary of methods and notation}

The possible CC0+DFT combinations presented here 
and their corresponding closed-shell
 energy expressions are summarized 
in Table \ref{tab:methods} (in this work, we deal only with 
closed shells).
All these methods have been implemented 
in a development version of \textsc{Gaussian}~\cite{Gaussian}.
A variant called CC0+$E_c^\text{DFT}$ is also introduced 
where the full $E_c^\text{DFA} [n_\uparrow, n_\downarrow ]$ is 
added to the CC0 energy. 
The purpose of including CC0+$E_c^\text{DFT}$ is to 
assess the effects of double counting, and how well 
do improved CC0+DFT combinations eliminate this problem. 
The explanation of the notation for the rest of the methods is as follows:
A ``p'' before the functional name indicates
that only parallel spin correlation from the DFA---and dRPA, for
the methods that include it---is added 
to CC0; the ``t'' that the full triplet-component 
of $T_2$ is added (see Section 2.2). 
The variant of SCAN called ``rSCAN'' uses the 
spin \textit{resolution} from Eq.~\ref{eq:eq20}, rather than 
that of Eq.~\ref{eq:Stoll}.
Finally, the ``LC-'' prefix (which stands for ``long-range
corrected'') specifies that the short- and long-range correlation terms
are evaluated with the DFA and the dRPA, respectively.
The long-range correction used here should not be confused 
with the one used in standard LC-KS-DFT; 
the former affects only the correlation, while the latter 
affects only the exchange. The exchange does not need 
corrections in CC0+DFT as it is calculated with the wavefunction 
method. 
We also remind the reader that the calculations are 
carried out in a non-self-consistent, post-CC0 manner; 
a self-consistent implementation is possible but 
not the focus of this exploratory paper.

\section{Results and discussion}

\subsection{Description of short-range dynamic correlation}

\begin{table*}
  \begin{center}
    \caption{Accurate~\cite{NIST,CCCBD}
      equilibrium distances ($R_e$, in $\AA$) 
      for a set of diatomics 
      and the deviations (calculated $-$ accurate, in 
      miliangstroms) from these distances for
      BD0 and BD0+DFT methods using a Cartesian 6-311++G(2df,2p) basis.
    ME is the mean error and MAE the mean absolute error.
    Results from CCD0 and CCD0+DFT are very similar.}
    \label{tab:bonds}
    \scalebox{0.8}{
      \begin{tabular*}{1.25\textwidth}{@{\extracolsep{\fill}}  ccccccccc  }
        \hline
                 &       &     & \multicolumn{6}{c}{BD0+} \\
                 \cline{4-9} 
                 Molecule & Accurate & BD0 & $E_c^\text{TPSS}$ & $E_c^\text{SCAN}$ & pTPSS & pSCAN & prSCAN & trSCAN \\
        \hline 
H$_2$ & 0.741 &  1 & -2 &  0 & 1  & 1  & 1  & 1 \\
LiH   & 1.596 &  6 & -10 & -12 & 3  & -2 & 1  & -5\\
HF    & 0.917 &  1 & -10 & -5 & -5 & -1 & -1 & -2\\
HCl   & 1.275 &  7 & -7 & -2 & 0  & 3  & 4  & 2 \\
Li$_2$& 2.673 & -28 & -28 & -50 & -2 & -20 & -11 & -28\\
C$_2$ & 1.243 &  4 & -9 & -6 & -1 & 0  & 0  & -2\\
N$_2$ & 1.098 &  1 & -8 & -5 & -2 & -1 & -1 & -2\\
F$_2$ & 1.412 &  15 & -34 & -26 & -14 & -6 & -2 & -11\\
Cl$_2$& 1.988 &  41 & -19 & -9 & 2  & 12  & 20  & 9 \\
ClF   & 1.628 &  25 & -19 & -8 & 0  & 8  & 12  & 5 \\
 \hline    
ME   &  & 7 & -15 & -12 & -2 & -1 & 2 & -3 \\
MAE  &  & 13 &  15 &  12 &  3 &  5 & 5 &  7 \\
\hline
      \end{tabular*}
    }
  \end{center}
\end{table*}

For testing the CC0+DFT description of 
short-range dynamic correlation, we consider the
equilibrium distances ($R_e$) and harmonic 
vibrational frequencies ($\omega_e$) of a set of ten first- 
and second-row diatomics that has been studied in previous 
MR+DFT works~\cite{Perez2007,Perez2007b}.
These molecules are listed in Table~\ref{tab:bonds}; 
they comprise well-known examples of single- and multiple-bonds, 
homonuclear and heteronuclear diatomics; 
accurate experimental data are available from 
established databases~\cite{NIST,CCCBD}. 
The motivation for studying $R_e$ and $\omega_e$ is as follows: 
$E_c^\text{DFA}$ increases in magnitude as the interatomic 
distance is reduced, increasing the bond strength. 
Because CC0+$E_c^\text{DFT}$ is deliberately constructed to 
have double counting from $E_c^\text{DFA}$, 
this approach should predict too short bond lengths 
and too high frequencies. 
If the rest of the CC0+DFT combinations in Table~\ref{tab:methods} 
really work, then these problems would disappear and 
improvement over CC0 should be observed.

The bond lengths predicted by BD0 and BD0+DFT 
are compared with accurate data in Table~\ref{tab:bonds}; 
results by CCD0 and CCD0+DFT are highly similar to those 
of their corresponding BD0 counterparts. 
Likewise, LC-CC0+DFT and CC0+DFT are not significantly different 
in this case. 
Note that the H$_2$ molecule, for which BD0 is exact, 
is included among the benchmark set.
The accurate data are from experiments~\cite{NIST,CCCBD}
and the difference
with BD0 for H$_2$ ($0.001 \, \AA$)
may be considered as an estimate of the effects
of basis set incompleteness (and, in principle, also the 
Born--Oppenheimer approximation).
As expected from the discussion above, 
BD0+$E_c^\text{DFT}$ consistently underestimates 
the bond lengths (mean error, ME = $-0.015 \, \AA$), 
while the BD0+DFT methods 
that avoid double counting do not do this and 
furthermore improve upon BD0:
the mean absolute errors (MAEs) of the different BD0+DFT combinations
are 2--4 times smaller than the MAE of BD0 ($0.013 \, \AA$), 
all providing very good results. 
Thus, the CC0+DFT methods appear to be working as intended in the
prediction of equilibrium distances.
Perhaps the only unexpected result is that there does not appear 
to be much difference between the different spin resolutions 
of SCAN (pSCAN and prSCAN), or between adding only parallel 
spin correlation or the full triplet component of $T_2$ 
(prSCAN and trSCAN). 
We also note that most of the largest errors occur in Li$_2$. 
This diatomic is somewhat challenging to describe because, 
even at equilibrium, there may be some static correlation 
present~\cite{Pollet2002}. 
Nevertheless, the errors for the $R_e$ of Li$_2$
obtained here are much smaller than those 
reported for MCSCF (error = 0.258 $\AA$) and 
MCSCF+DFT (error $\approx$ 0.150 $\AA$) 
in Ref.~\cite{Perez2007}.

General observations are mostly similar for the 
harmonic vibrational frequencies shown in Table~\ref{tab:freqs}.
Again, we exclude LC-CC0+DFT data because
the effect of the long-range correction with dRPA correlation 
is negligible on the calculated frequencies.
A difference though is that there is more variation between 
BD0+DFT and CCD0+DFT, and hence we include data for both
in this Table. 
The effect of double counting is as expected 
and CC0+$E_c^\text{DFT}$ consistently overestimates the frequencies
(ME = MAE $\approx$ 70--80 cm$^{-1}$ with TPSS, 
 40--60 cm$^{-1}$ with SCAN). 
Results from CCD0 and BD0 are close to each other (MAE = 
22 cm$^{-1}$ for both), but adding pTPSS to the former 
worsens results (MAE = 42 cm$^{-1}$) while it does not 
for the latter (MAE = 25 cm$^{-1}$). 
Adding pSCAN, prSCAN, or trSCAN to CCD0 gives similar 
or slightly worse results than CCD0; 
for BD0, predictions become similar or better. 
The densities from the approximate Brueckner orbitals may therefore be more 
adequate inputs for $E_c^\text{DFA}$ than the RHF densities. 
This observation may be related to the assertions about the 
physical relevance of Brueckner orbitals---connections to 
DFT~\cite{Scuseria1995}
and even the Kohn--Sham orbitals~\cite{Lindgren2002}
have been suggested in the literature. 
Note also that the exact Brueckner determinant can be defined 
as the Slater determinant having the largest overlap with FCI, 
and thus the densities can be presumed to be of better quality 
than those from RHF.
Overall, all of the combinations of CCD0 or BD0 with SCAN 
give satisfactory results for both bond lengths and frequencies; 
mixtures with TPSS are also reliable for these properties 
except for CCD0+pTPSS, which tends to overestimate the frequencies.

\begin{table*}
  \begin{center}
    \caption{Accurate~\cite{NIST,CCCBD} harmonic vibrational
      frequencies ($\omega_e$, in cm$^{-1}$) 
      for a set of diatomics 
      and the deviations (calculated $-$ exact) from these values for
      CC0 and CC0+DFT methods using a Cartesian 6-311++G(2df,2p) basis.
    ME is the mean error and MAE the mean absolute error.}
    \label{tab:freqs}
    \scalebox{0.8}{
      \begin{tabular*}{1.25\textwidth}{@{\extracolsep{\fill}}  ccccccccc  }
        \hline
                 &       &     & \multicolumn{6}{c}{CCD0+/BD0+}\\
                 \cline{4-9}  
                 Molecule & Accurate & CCD0/BD0 & $E_c^\text{TPSS}$ & $E_c^\text{SCAN}$ & pTPSS & pSCAN & prSCAN & trSCAN \\
        \hline 
H$_2$ & 4401 & 23/11   & 65/60   & 42/23  & 23/11  & 23/11   & 23/11   & 23/11 \\
LiH   & 1406 & -6/-3   & 32/30   & 39/31  & 0/4    & 0/3     & -3/-11  & 9/0   \\
HF    & 4138 & 36/19   & 160/156 & 103/84 & 102/84 & 22/24   & 44/24   & 58/38  \\
HCl   & 2991 & -38/-38 & 70/65   & 35/21  & 7/21   & -22/-21 & -32/-47 & -17/-29 \\
Li$_2$& 351  & -3/32   & 19/18   & 11/10  & 9/7    & 9/26    & 2/16    & 12/5   \\
C$_2$ & 1855 & 22/12   & 106/87  & 77/58  & 89/51  & 40/29   & 40/26   & 50/36  \\
N$_2$ & 2359 & 36/19   & 108/87  & 77/54  & 63/40  & 37/25   & 44/21   & 51/29  \\
F$_2$ & 917  & 17/-31  & 132/87  & 110/65 & 81/8   & 64/9    & 52/0    & 76/22  \\
Cl$_2$& 560  & -19/-19 & 45/41   & 32/27  & 22/17  & 7/10    & -2/-6   & 10/5   \\
ClF   & 786  & -24/-34 & 54/41   & 36/18  & 20/6   & 7/0     & -4/-15  & 12/-4  \\
 \hline    
ME   &  & 4/-3 & 79/67  & 56/39 &  42/25 & 19/11 & 16/2 &  28/11 \\
MAE  &  & 22/22 &  79/67 & 56/39 & 42/25 & 23/16 & 25/18 & 32/18  \\
\hline
      \end{tabular*}
    }
  \end{center}
\end{table*}

As noted above, it is rather surprising that different 
spin resolutions, or whether one uses pDFT or tDFT, have 
little effect on the calculated bond lengths and frequencies. 
However, these properties depend on relative energies only; 
total energies are certain to be affected by factors such as 
the incorporation of parallel spin \textit{vs.} triplet pairing
channel correlation (\textit{i.e.}, pDFT \textit{vs.} tDFT). 
It is then logical to ask which of the different flavors of 
CC0+DFT provides more accurate total energies. 
Table \ref{tab:neon} compares the accurate~\cite{Zhao2008}
energy for the Neon 
atom with BD0 and BD0+DFT estimates; 
results from CCD0 and CCD0+DFT differ by less than 
1 milihartree from their BD0 analogues.
The correlation missed by BD0 in Ne is substantial (120 
milihartrees). If raw $E_c^\text{DFT}[n_\uparrow, n_\downarrow ]$ is added, 
results are worsened due to overcorrelation (error $\approx$ $-$225
milihartrees). In contrast, adding only parallel spin 
correlation improves the total energy, although 
underestimating it (errors = 41 and 20 milihartrees for pTPSS and 
pSCAN, respectively). 
The best agreement with the accurate energy is obtained 
when the correlation from the full triplet pairing channel is 
added to BD0 and using the spin resolution of Eq.~\ref{eq:eq20}
(errors = 2 and $-$10 milihartrees for tTPSS, and trSCAN, 
respectively), but there is overcorrelation if the 
spin resolution of Stoll \textit{et al.}~\cite{Stoll1978} is used with SCAN 
(error = $-$29 milihartrees for tSCAN). 
These observations are in
line with the theoretical arguments from Sections 2.2 and 2.4; 
\textit{i.e.}, that BD0 misses the full triplet component 
of $T_2$, and that the resolution by Stoll \textit{et al.}~\cite{Stoll1978}
exaggerates same-spin correlation in most regimes 
of the uniform electron gas.

\begin{table*}
  \begin{center}
    \caption{Accurate~\cite{Zhao2008} total
      energy for the Neon atom (Hartrees) and deviations 
      (calculated $-$ exact, in milihartrees) from this value for BD0 and BD0+DFT methods using 
      a Cartesian cc-pwCV5Z basis. Results from CCD0 and CCD0+DFT differ by less than 
    1 milihartree from their BD0 counterparts.}
    \label{tab:neon}
    \scalebox{0.8}{
      \begin{tabular*}{1.25\textwidth}{@{\extracolsep{\fill}}  cccccccccc  }
        \hline
                 &       &     \multicolumn{8}{c}{BD0+} \\
                 \cline{3-10} 
                 Accurate & BD0 & $E_c^\text{TPSS}$ & $E_c^\text{SCAN}$ & pTPSS & tTPSS & pSCAN & tSCAN & prSCAN & trSCAN \\
                 \hline
                 -128.938 & 120 & -229 & -224 & 41 & 2 & 20 & -29 & 33 & -10 \\
        \hline 
      \end{tabular*}
    }
  \end{center}
\end{table*}

\begin{table*}
  \begin{center}
    \caption{Accurate
      proton affinities (PA = $E(M) - E(MH^+)$, in kcal/mol) for eight 
      molecules and deviations 
      (calculated $-$ exact, in kcal/mol) 
      from these values for TPSS, SCAN, BD0, and BD0+DFT using 
      a Cartesian MG3S basis. 
      The TPSS and accurate data are from Ref.~\cite{Zhao2006}.
      Results from CCD0 and CCD0+DFT are very similar 
      to those from  their BD0 counterparts.}
    \label{tab:pa8}
    \scalebox{0.77}{
      \begin{tabular*}{1.3\textwidth}{@{\extracolsep{\fill}}  ccccccccccc  }
        \hline
                 &    &   &  & &   \multicolumn{6}{c}{BD0+} \\
                 \cline{6-11} 
                 Molecule & Accurate & TPSS & SCAN & BD0 & pTPSS & tTPSS & pSCAN & tSCAN & prSCAN & trSCAN \\
                 \hline
                 NH$_3$     &  211.90 & 1.70 & 0.72  & 2.84  & 3.53  & 3.88  & 1.22  & 0.41  & 1.70  & 1.13  \\ 
                 H$_2$O     &  171.80 & 0.40 & 0.24  & 1.44  & 2.43  & 2.93  & 0.39  & -0.14 & 0.69  & 0.31  \\ 
                 C$_2$H$_2$ &  156.60 & 4.60 & 3.83  & 2.89  & 4.01  & 4.56  & 0.95  & -0.02 & 1.30  & 0.50  \\ 
                 SiH$_4$    &  156.50 & 3.10 & -0.34 & 2.75  & 0.46  & -0.68 & 0.24  & -1.01 & 0.84  & -0.12 \\ 
                 PH$_3$     &  193.10 & 2.60 & -0.02 & 4.55  & 1.04  & -0.71 & -0.26 & -2.66 & 1.09  & -0.65 \\ 
                 H$_2$S     &  173.70 & 3.30 & 1.45  & 3.72  & 2.74  & 2.24  & 1.04  & -0.30 & 1.79  & 0.83  \\ 
                 HCl        &  137.10 & 3.10 & 1.50  & 2.92  & 3.18  & 3.31  & 1.42  & 0.68  & 1.83  & 1.29   \\ 
                 H$_2$      &  105.90 & 2.40 & -0.01 & -0.44 & -0.44 & -0.44 & -0.44 & -0.44 & -0.44 & -0.44 \\ 
        \hline 
        ME  & & 2.65 & 0.92 & 2.58 & 2.12 & 1.89 & 0.57 & -0.43 & 1.11 & 0.36 \\
        MAE & & 2.65 & 1.01 & 2.69 & 2.23 & 2.34 & 0.75 & 0.71  & 1.22 & 0.66 \\
        \hline
      \end{tabular*}
    }
  \end{center}
\end{table*}

Another test for dynamic correlation 
based on relative energies, but more sensitive 
than the bond lengths and frequencies above,
is the description of proton affinities. 
When a proton is added to a neutral molecule, 
the density is redistributed and dynamic 
correlation changes subtly. This effect 
is reasonably well described by density functionals, 
which have average errors of about 1--3 kcal/mol in small 
molecules~\cite{Zhao2006,Staroverov2003}. 
Table~\ref{tab:pa8} compares the experimental (ZPE-corrected)
proton affinities of eight molecules
with those calculated by TPSS, SCAN, BD0, and BD0+DFT.
The geometries (which are optimized at the  MP2/6-31G(2df,p) level) 
and reference data were taken from Ref.~\cite{Zhao2006};
the effects of geometric relaxation are known to be 
negligible~\cite{Staroverov2003}.
Again, we focus on BD0+DFT methods because CCD0+DFT results 
are very similar and the long-range correction of 
LC-BD0+DFT has little effect on the proton affinities. 
Pure BD0 tends to overestimate the proton affinities 
(ME = 2.58 kcal/mol) and has 
the largest MAE of all methods (2.69 kcal/mol). 
TPSS results are similar (ME = MAE = 2.65 kcal/mol), 
but SCAN is considerably better (MAE = 1.01 kcal/mol).
Previous works noted that TPSS did not improve upon 
its Perdew--Kurth--Zupan--Blaha~\cite{Perdew1999}
predecessor in the 
prediction of proton affinities~\cite{Staroverov2003}.
The fact that SCAN is better than TPSS 
at describing the effect of protonization is 
reflected in BD0+DFT: the BD0+TPSS errors
(MAE $\approx$ 2.3 kcal/mol) are 2--3 times larger than the BD0+SCAN errors. 
The best results are obtained when the full triplet-pairing 
component of $T_2$ is added to BD0 using the spin resolution 
of Section 2.4 (BD0+trSCAN, MAE = 0.66 kcal/mol). 
The observations here point toward a relationship between 
the accuracies of the DFA and BD0+DFA for a given property. 
Note, however, that not any functional can be utilized in 
CC0+DFT. For example, the Lee--Yang--Parr~\cite{Lee1988}
(LYP) functional models all correlation as being 
opposite-spin~\cite{Filatov2005}, so that there is no parallel-spin 
correlation that can be extracted from LYP.

\subsection{Description of long-range dynamic correlation}

\begin{figure}
  \centering
  \includegraphics[width=0.46\textwidth]{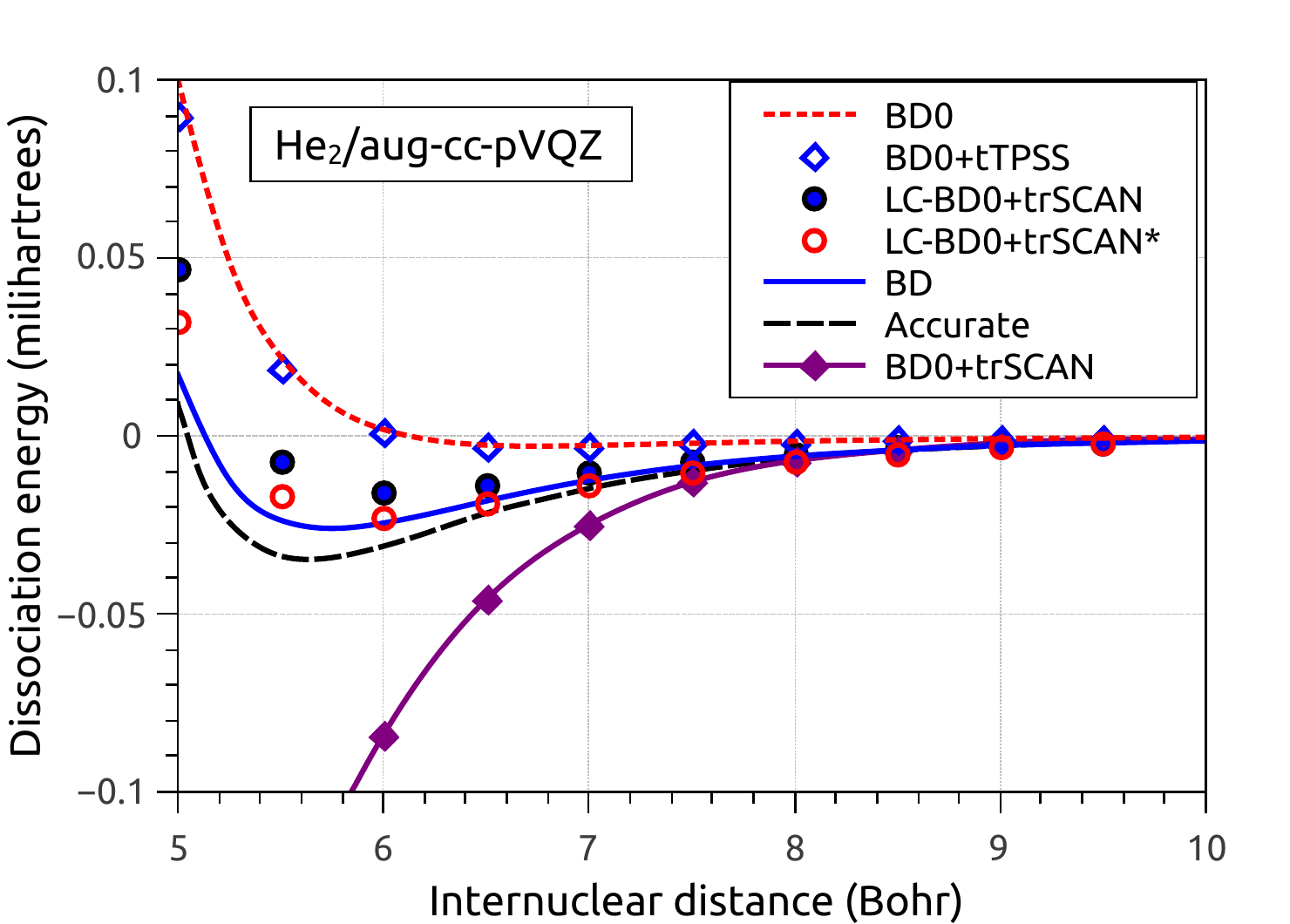}
  \caption{Counterpoise-corrected dissociation energy profiles for 
    the Helium dimer calculated by various methods using 
    a Cartesian aug-cc-pVQZ basis. LC-BD0+trSCAN uses 
    $\mu = 0.4$ au.; BD0+tSCAN has the same behavior as 
    BD0+trSCAN.
    The curve marked as LC-BD0+trSCAN* uses the 
    semiempirical prefactor of 1.5 on the dRPA correlation 
    of Ref.~\cite{Janesko2009}. 
    The accurate data are from  Ref.~\cite{Tang2003}. }
\label{fig:he2}
\end{figure}

In all the benchmarks that we have hitherto discussed,
the use of LC-CC0+DFT over CC0+DFT has negligible impact on 
the results. 
However, the effect of the long-range, dRPA-based, correction 
can be dramatic on the description of van der Waals interactions. 
One such example, the dissociation of a Helium dimer, 
is shown in Figure~\ref{fig:he2}.
The accurate data shown there  has been taken from 
Ref.~\cite{Tang2003}.
Standard Brueckner doubles (BD) correctly describes the dissociation 
profile of He$_2$. The correlation present in BD but not in BD0
is crucial in this case; the latter is repulsive in the region 
near the correct minimum. 
The reason for the failure of BD0 for long-range interactions 
can be explained as follows: 
BD0 misses equal-spin correlation (plus an
identical-in-magnitude $m = 0$ component of
$T_2^{[1]}$) which, in the short-range, is generally much smaller than the 
opposite-spin contributions because same-spin electrons avoid
each other due to the Pauli principle (antisymmetry of the wavefunction).
In the long-range, however, electrons are far apart from each other 
and their spin is no longer relevant on their dynamic correlation. 
Hence, the correlation missing in BD0 becomes more important 
in the long range.

From Figure~\ref{fig:he2} we also see that BD0+TPSS adds 
a negligible contribution to the BD0 description of the He$_2$ 
dissociation; the same-spin TPSS correlation is practically
zero along the whole curve. 
Surprisingly, BD+SCAN is radically different from 
BD0+TPSS, but fails badly in an opposite way:
SCAN exaggerates the long-range parallel-spin correlation 
and leads to far too much binding. 
This wrong behavior is corrected when the DFA 
correlation is evaluated in the short-range only and complemented 
with dRPA correlation in the long range. 
The LC-BD0+trSCAN and BD curves are close to each other, 
indicating that adding the full triplet component of 
$T_2$ via our SR-SCAN/LR-dRPA model correctly incorporates 
the correlation that is missing in BD0. 
It is worth pointing out that 
Janesko \textit{et al.}~\cite{Janesko2009} found
that using an empirical scaling of 1.5 on the dRPA correlation improves
the agreement of LC-LDA+dRPA with accurate data for noble gas dimers, and 
that here too we observe improvement if this factor 
is used (LC-BD0+trSCAN* curve). 
Note also that we do not show data for all possible method combinations 
in Figure~\ref{fig:he2}, but the performance of the techniques
not shown can be inferred by the reader due to the fact that
many BD0+DFT variants differ only by a multiplicative factor on 
the correlation added to BD0. 
For example, the
LC-BD0+prSCAN curve (not shown in Figure~\ref{fig:he2})
would be qualitatively similar to the LC-BD0+trSCAN one, 
but less accurate in quantitative terms.

\begin{table}
  \begin{center}
    \caption{Accurate binding energy for the Neon dimer 
    compared with counterpoise-corrected calculations by BD0 and 
    BD0+DFT methods 
    at the experimental equilibrium distance of 5.841 au
    using a Cartesian aug-cc-pVQZ basis.
    The geometries and accurate binding energy were taken 
    from Ref.~\cite{Zhao2005}.
    All energies are in kcal/mol.}
    \label{tab:Ne2}
    \scalebox{0.9}{
      \begin{tabular*}{0.49\textwidth}{@{\extracolsep{\fill}}  lc  }
        \hline
        Method & Binding Energy \\
        \hline
        Accurate & 0.08 \\
        BD0 & -0.03 \\
        BD0+pTPSS & -0.03 \\
        BD0+tTPSS & -0.02 \\
        BD0+pSCAN & 0.03 \\
        BD0+tSCAN & 0.06 \\
        BD0+prSCAN & 0.01 \\
        BD0+trSCAN & 0.03 \\
        LC-BD0+pTPSS & 0.00 \\
        LC-BD0+tTPSS & 0.02 \\
        LC-BD0+prSCAN & 0.01 \\
        LC-BD0+trSCAN & 0.03 \\
        LC-BD0+tTPSS$^*$ & 0.05 \\
        LC-BD0+trSCAN$^*$ & 0.05 \\
        \hline    
        \multicolumn{2}{l}{\footnotesize{$^*$Using the semiempirical 
        prefactor of 1.5 on the}}\\
          \multicolumn{2}{l}{\footnotesize{dRPA correlation 
        of determined on Ref.~\cite{Janesko2009}.}}
      \end{tabular*}
    }
  \end{center}
\end{table}

On Table~\ref{tab:Ne2} the binding energies of the Neon dimer
calculated by BD and BD0+DFT at the experimental
equilibrium distance of 5.841 au are compared with accurate 
estimates from standard databases~\cite{Zhao2005}.
Once more, BD0 fails to bind the noble gas dimer, 
and so do BD0+pTPSS and BD0+tTPSS. 
Remarkably though, all combinations of BD0 with SCAN bind 
Ne$_2$, showing best agreement with the reference energy
when the full triplet-pairing component of $T_2$ is included.
When the long-range DFA correlation is replaced by dRPA correlation, 
variants using TPSS and SCAN 
all bind the dimer, but tSCAN-based methods are most accurate.
LC-BD0+pTPSS provides the smallest binding energy (0.004 kcal/mol)
and LC-BD0+trSCAN the largest one (0.03 kcal/mol).
As occurs for He$_2$, 
if the semiempirical factor of 1.5 determined on Ref.~\cite{Janesko2009} 
is used to scale $E_c^\text{dRPA}$ the agreement between 
LC-BD0+tDFT and experiment is improved.

\begin{table}
  \begin{center}
    \caption{Accurate binding energies for 
      ethylene (C$_2$H$_4$) and acetylene (C$_2$H$_2$) dimers
      compared with counterpoise-corrected calculations by BD0 and 
      BD0+DFT methods using a Cartesian MG3S basis. 
      The geometries and accurate data were taken 
      from Ref.~\cite{Zhao2005}. 
      All energies are in kcal/mol.}
    \label{tab:pi}
    \scalebox{0.9}{
      \begin{tabular*}{0.49\textwidth}{@{\extracolsep{\fill}}  lcc  }
        \hline
        & \multicolumn{2}{c}{Binding Energy} \\
        \cline{2-3}
        Method & (C$_2$H$_2$)$_2$ & (C$_2$H$_4$)$_2$ \\
        \hline
        Accurate & 1.34 & 1.42  \\
        BD0 & 0.38 & -0.42 \\
        BD0+pTPSS & 0.64 & 0.07 \\
        BD0+tTPSS & 0.76 & 0.31 \\
        BD0+pSCAN & 1.13 & 1.03 \\
        BD0+tSCAN & 1.50 & 1.76 \\
        BD0+prSCAN & 0.95 & 0.23 \\
        BD0+trSCAN & 1.24 & 0.56 \\
        \hline    
      \end{tabular*}
    }
  \end{center}
\end{table}

It should also be mentioned that there are some types of weak 
interactions that are captured by semilocal functionals, 
as these can describe the intermediate-range part of 
the van der Waals interaction~\cite{Sun2015,Tao2010,Sun2013}. 
As an example, we consider the $\pi$-$\pi$ interactions 
in ethylene and acetylene dimers with geometries 
from a standard dataset~\cite{Zhao2005}. 
Accurate data for the binding energies of these species 
are compared with BD0 and BD0+DFT predictions on Table~\ref{tab:pi}. 
At the accurate equilibrium distance, BD0 significantly underestimates 
the binding energies for both dimers and predicts only the acetylene 
dimer to be bound. 
All of the BD0  combinations with DFT 
improve the former's 
results and bind both dimers, 
but the best agreement with the reference data 
is given by BD0+tSCAN and BD0+trSCAN. 
It was noted on Ref.~\cite{Sun2015} that SCAN was more 
accurate than TPSS for describing weak interactions; 
a consequence, most likely, of the appropriate norming of SCAN. 
Hence, we observe again a correlation between 
the adequacy of the DFA for a given property,
and that of BD0+DFA methods for the same type 
of calculation.

In this Section, we have focused on BD0+DFT results; 
CCD0+DFT results have been omitted. The reason for this is 
that, in our experience,
BD0+DFT and CCD0+DFT are largely similar when describing 
problems dominated by dynamic correlation (with BD0 being slightly 
better). 
Thus, the observations for CCD0-based methods are analogous to those
of their BD0 counterparts and discussion about the former has 
been left out to avoid repetitiveness and long-windedness. 
Nevertheless, as we discuss next,
there can be important differences between 
CCD0 and BD0 when static correlation is involved.

\subsection{Simultaneous description of static and 
dynamic correlation}

\begin{figure*}
  \centering
  \includegraphics[width=0.46\textwidth]{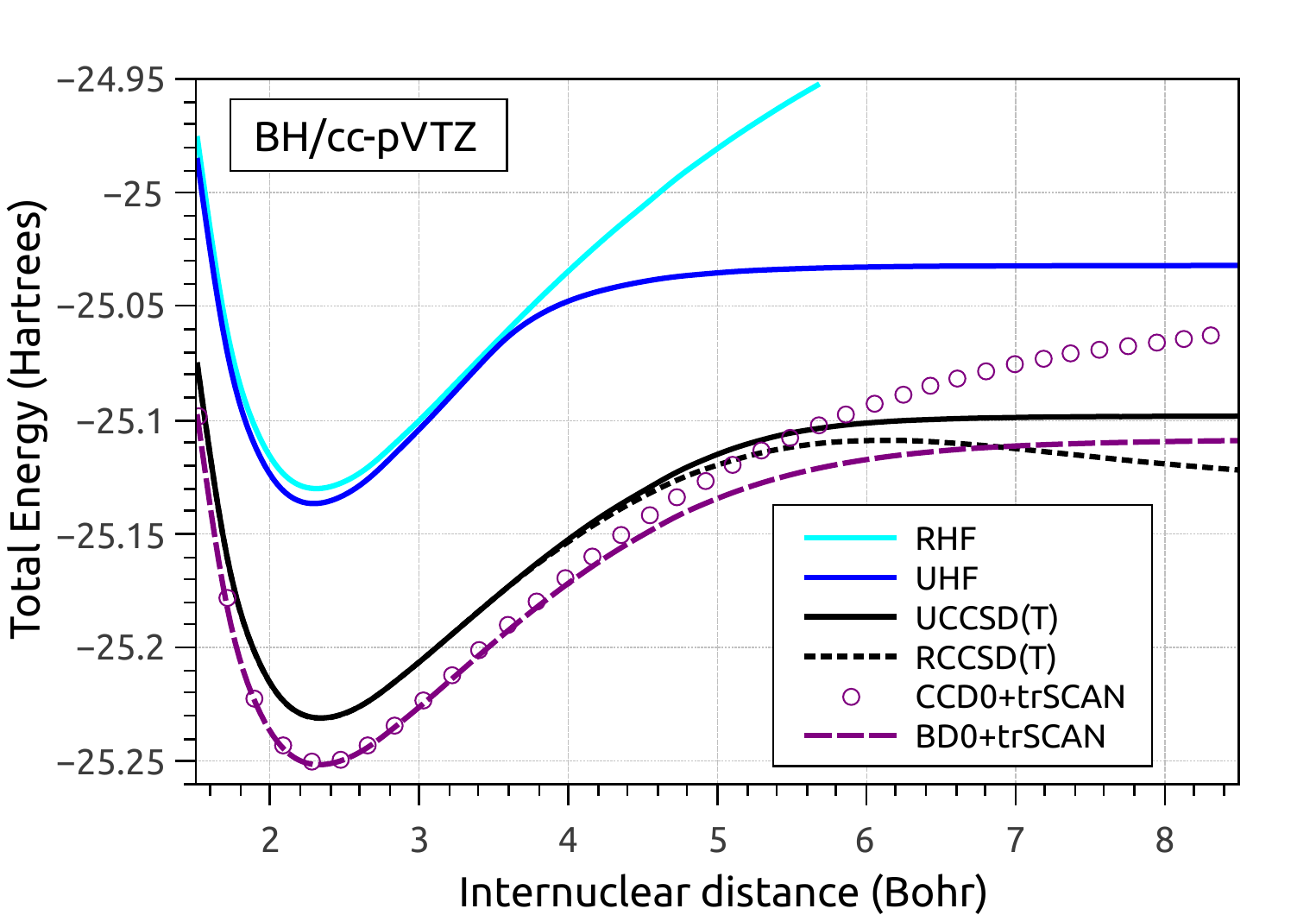}
  \includegraphics[width=0.46\textwidth]{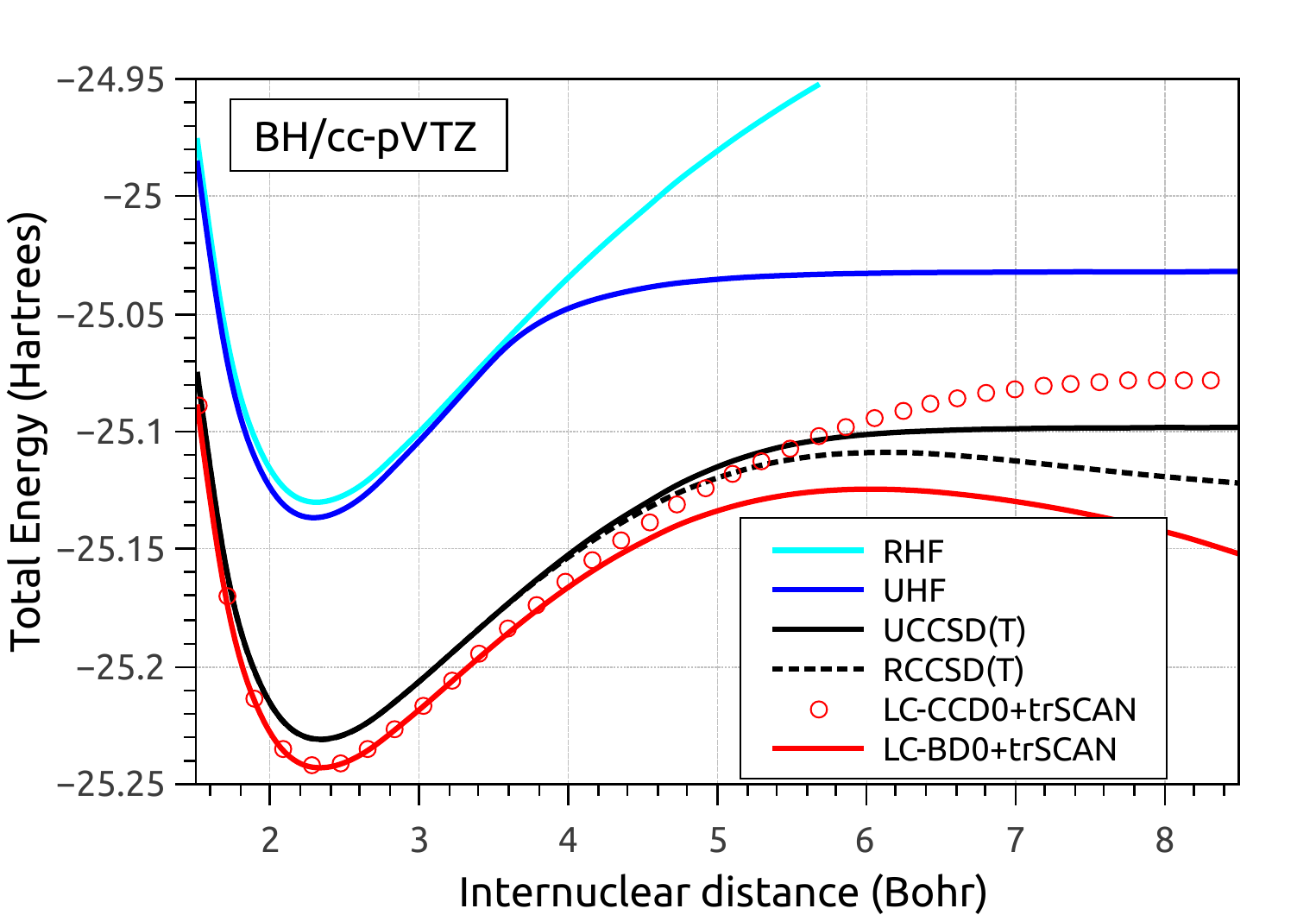}
  \caption{Dissociation energy profiles for 
    boron hydride calculated by various methods using 
    a Cartesian cc-pVTZ basis.}
\label{fig:bh}
\end{figure*}

Molecular dissociations are one of the paradigms of static 
correlation in quantum chemistry. 
Figure~\ref{fig:bh} shows the dissociation energy curves for 
boron hydride calculated by various standard and CC0+rSCAN 
methods; CC0+TPSS and CC0+SCAN 
variants are similar but slightly higher and lower, 
respectively, in energy than CC0+rSCAN. 
The UCCSD(T) (unrestricted CCSD with perturbative triples)
data in this Figure
may be considered as an accurate reference, 
as this technique is considerably parallel to FCI for this 
system~\cite{Dutta2003}
(nonparallelity error $\approx$ 2 kcal/mol).
Note, however, that restricted CCSD(T) experiences a breakdown at
B--H distances greater than about 5.5 au. 
As noted in the previous sections, CCD0- and BD0-based methods
are highly similar when the problem is dominated by dynamic correlation 
(in this case, near equilibrium, although the existence of 
a UHF solution slightly lower in energy than RHF suggests the presence of
some static correlation too; similar observations 
have been noted by Pollet \textit{et al.}~\cite{Pollet2002}). 
However, this is no longer true at large bond lengths, where
CCD0 becomes significantly higher in energy than BD0. 
In general, the BD0 description of bond breaking is significantly better 
than that of CCD0. 
This is easily understood by considering the dissociation of 
H$_2$: BD0 is exact but CCD0 is not because, 
in order to be exact for two-electron systems, singles contributions 
must be included either explicitly or via
the Brueckner orbitals.
As the total correlation increases when the bond is extended
beyond the Coulson--Fischer point, the singles contributions 
that are missing in CCD0 become more important. 
This analysis extends to other chemical bonds in general.

\begin{figure}
  \centering
  \includegraphics[width=0.46\textwidth]{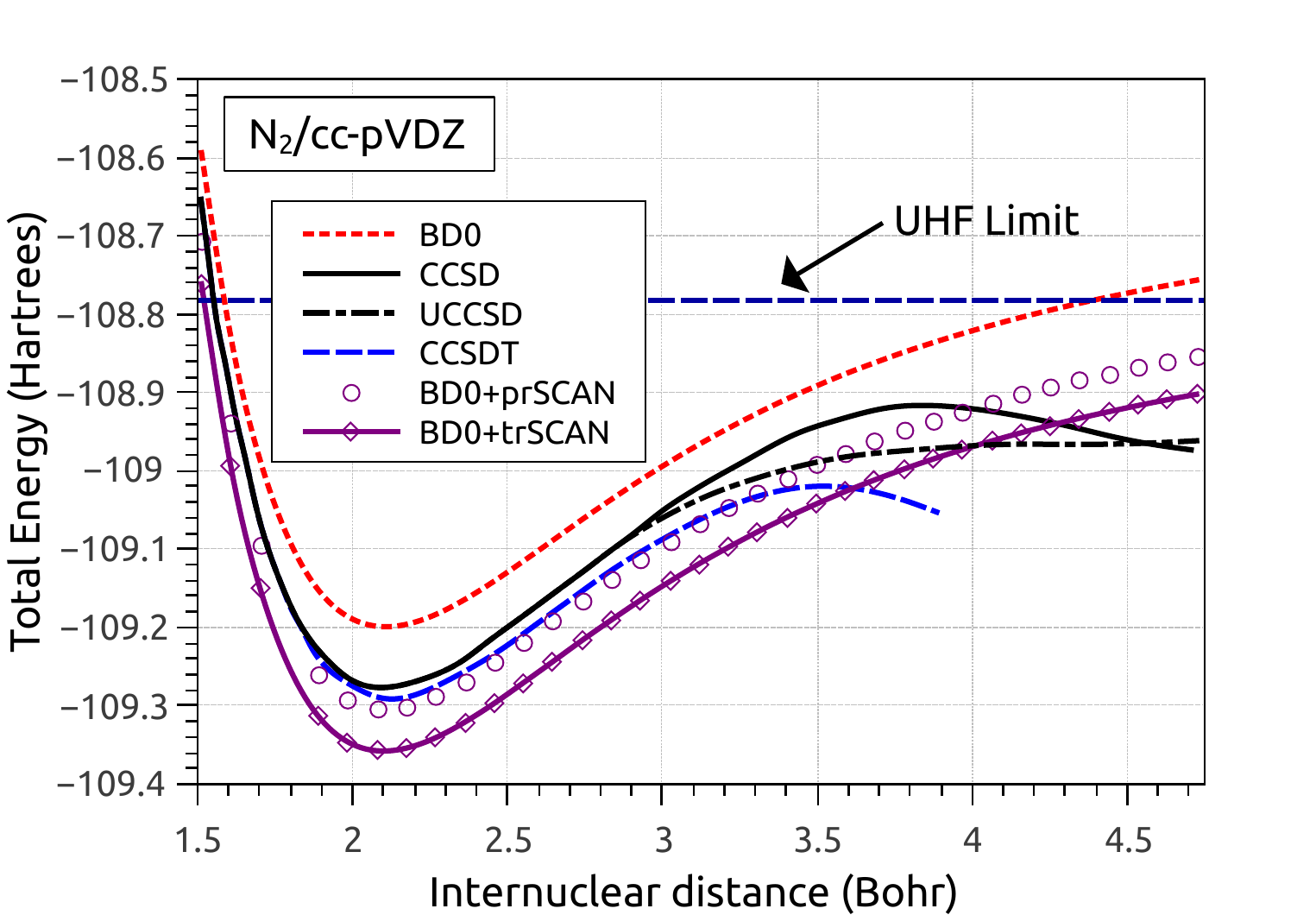}
  \includegraphics[width=0.46\textwidth]{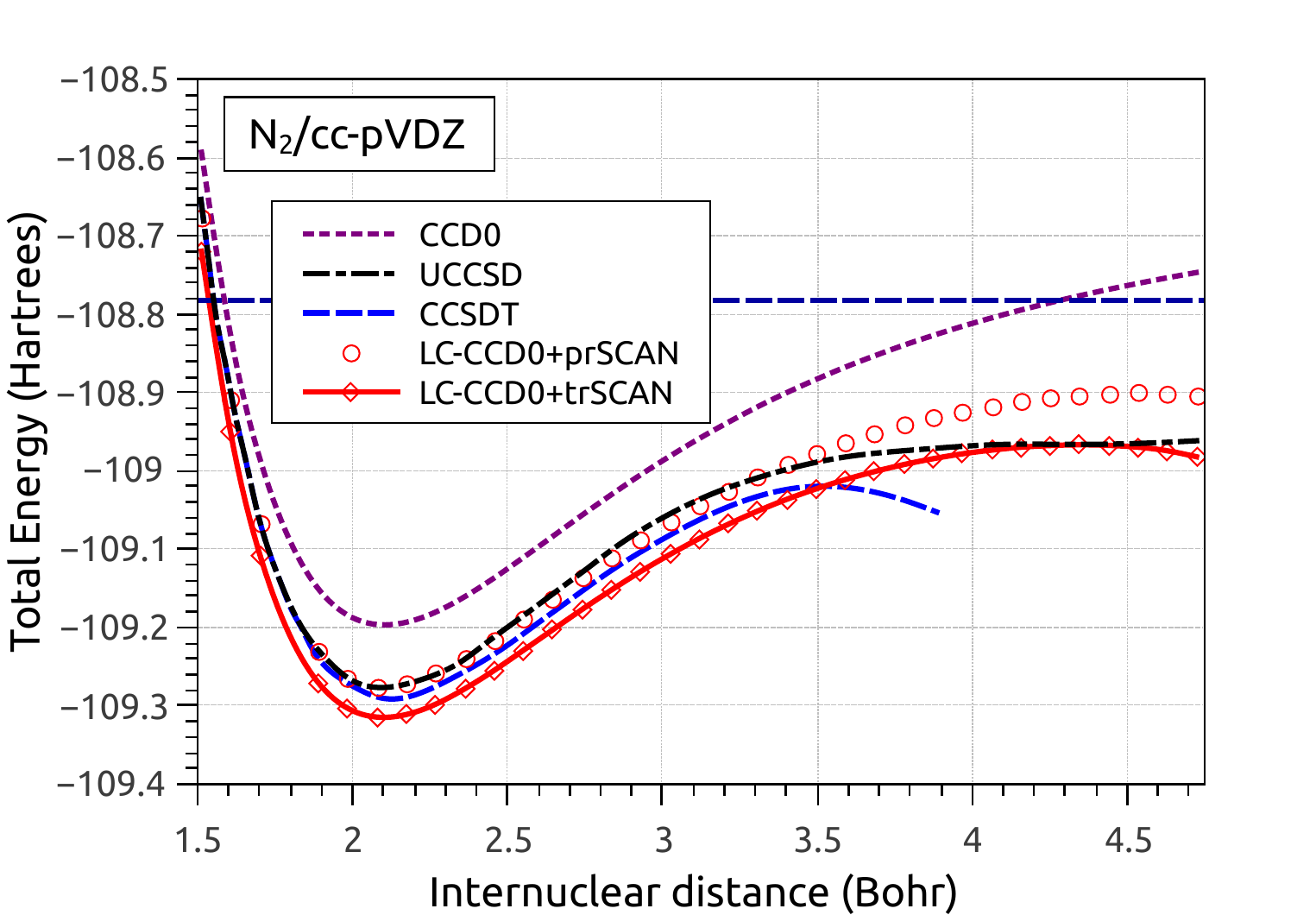}
  \caption{Dissociation energy profiles for 
    the N$_2$ molecule calculated by various methods using 
    a Cartesian cc-pVDZ basis.}
\label{fig:n2}
\end{figure}

Although the BD0+trSCAN curve in Fig.~\ref{fig:bh} provides 
a very good description of the BH dissociation (and without
resorting to symmetry breaking, like UCCSD(T)),
LC-BD0+trSCAN exhibits wrong behavior at 
large bond lengths. 
The dRPA correlation increases too much beyond 
$R_\text{B--H} \approx 6$ au, resulting in a ``bump'', 
similar to that of RCCSD(T), characteristic of common
perturbative and RPA-based approaches. 
In the case of LC-CCD0+trSCAN, the substantial dRPA correlation 
actually helps to correct, to a certain extent, 
the too-high dissociation limit of CCD0+trSCAN. 
When breaking multiple bonds, coupled cluster methods for 
strong correlation---CCD0, BD0, and pCCD---all tend to a 
too-high energy limit (sometimes higher than UHF, \textit{e.g.}, 
in N$_2$; see Figure~\ref{fig:n2}). 
Thus, adding dRPA correlation in a manner similar to the one 
done here may provide a route to alleviate this 
problem of singlet-paired coupled cluster techniques.
This, however, requires further investigation that is beyond 
the scope of the present paper. 
In the case of the dissociation of N$_2$ shown in Figure~\ref{fig:n2}, 
there also appears to be an improvement of the too-high 
dissociation energy limit of CCD0 by the long-range dRPA correlation. 
LC-CCD0+trSCAN is quite close energetically to UCCSD and 
binding energies are no longer exaggerated. 
The BD0+DFT curves are above LC-CCD0+DFT at large 
internuclear distances, but LC-BD0+DFT (not shown) leads to too 
much correlation at dissociation and a pronounced bump similar
to that seen in boron hydride. 
Note also that although this is one 
of the cases in which BD0 goes to a too high limit, 
CCSD and CCSDT fail very badly while BD0+DFT provides 
a reasonable description of the dissociation 
(and is also accurate near equilibrium, as evidenced 
by the data in Tables~\ref{tab:bonds} and \ref{tab:freqs}).

\begin{figure}
  \centering
  \includegraphics[width=0.16\textwidth]{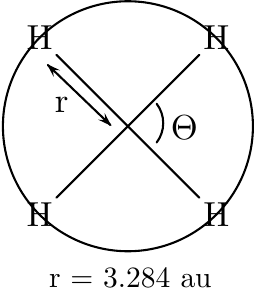}
  \includegraphics[width=0.46\textwidth]{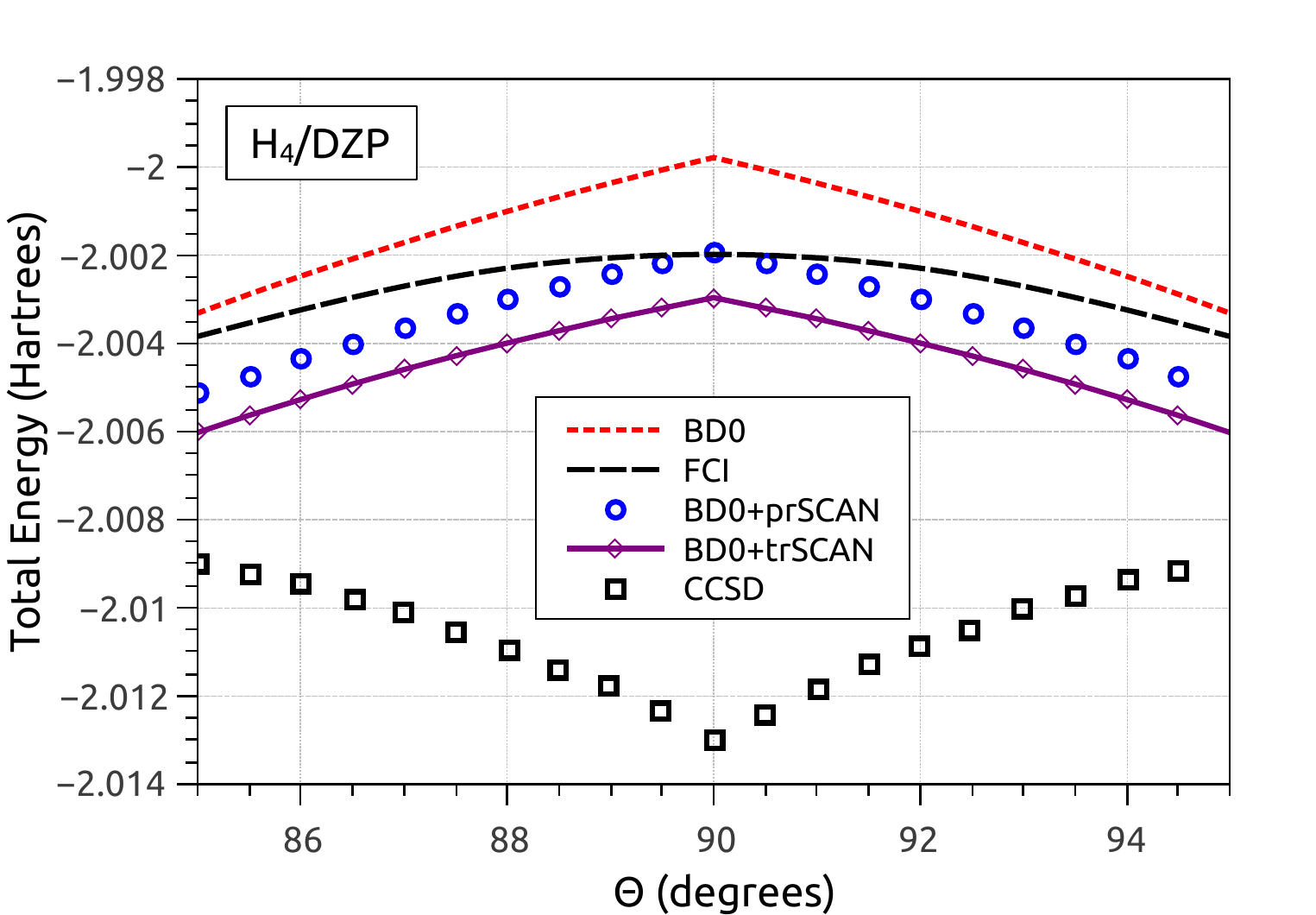}
  \caption{Geometry of H$_4$ on a circle with a radius of 3.284 au. 
    Dependence of the total energy on the angle $\Theta$ for 
    various methods using a Dunning DZP basis.
  The FCI data are from Ref.~\cite{VanVoorhis2000}. }
\label{fig:h4}
\end{figure}

Another problem for which traditional coupled cluster fails 
badly is the deformation of H$_4$ from a rectangular to a
square geometry~\cite{Bulik2015}.
Figure~\ref{fig:h4} shows this model (which has been studied 
extensively~\cite{Paldus1993,Kowalski1998a,Kowalski1998b,%
Kowalski1999,VanVoorhis2000})
schematically. 
The H atoms are confined to a circle of radius 3.284 au
and preserve $D_{2h}$ symmetry, 
so that the geometry depends on a single parameter:
the angle $\Theta$ that bisects two lines connecting
opposing H atoms. 
This Figure also shows the dependence of 
the energy on $\Theta$ for various methods, including FCI
data from Ref.~\cite{VanVoorhis2000}. 
It is seen that CCSD predicts a trend that is opposite 
to that of FCI, with a minimum at $\Theta = 90^\circ$ (square geometry),
where it overcorrelates the most. 
In contrast, BD0 and BD0+DFT methods have good qualitative and 
quantitative agreement with FCI, albeit they exhibit a 
discontinuity in the first derivative of the energy 
with respect to $\Theta$ at $\Theta = 90^\circ$ that is not 
present in the FCI curve. 
Of the BD0+DFT methods, only BD0+rSCAN data are shown in
Figure~\ref{fig:h4}, but using TPSS gives very similar results,
while BD0+SCAN yields slightly lower total energies. 

A benchmark for which KS-DFT methods fail catastrophically 
is the Beryllium isoelectronic series (\emph{i.e.}, 
$X^{(Z-4)+}$ ions).
In this system, the angular $s^2 \to p^2$ static correlation
increases linearly with increasing nuclear charge $Z$;
this effect is poorly described by common DFAs (even if one 
tries to break spin symmetry, because the $X^{(Z-4)+}$ ions are 
RHF $\to$ UHF stable). 
Table~\ref{tab:beiso} compares the accurate 
energies~\cite{Petersson1981} for this series
with energies from TPSS, SCAN, BD0, and their combinations; 
CCD0+DFT and BD0+DFT results are not significantly different in this 
case.
The nonparallelity errors (NPE)---the difference between the maximum and 
minimum errors---are also provided as an estimate of
how well is the linear trend of the correlation described. 
As expected, TPSS and SCAN fail badly to capture this trend with 
NPEs of 40.6 and 87.9 milihartrees, respectively. 
Much better results are given by BD0 (NPE = 5.2 milihartrees), 
although there is underestimation of the energy due to the missing 
same-spin correlation (ME = MAE = 7.1 milihartrees). 
All of the BD0+DFT combinations capture correctly the dependence 
of the energy with respect to $Z$, and are similar or better than BD0
with NPEs in the range of 2.0--7.3 milihartrees. 
The best results are given by BD0+pTPSS (ME, MAE, and NPE of 
$-$2.5, 2.5, and 1.2 milihartrees, respectively). 
The BD0+DFT methods tend to overestimate the correlation, 
which may be a result of the approximate functional and/or of the 
resolution of the equal-spin correlation. Indeed, the latter 
appears to be a factor because BD0+SCAN overcorrelates more 
than BD0+rSCAN. 
It is also possible that the overcorrelation in +tDFT methods 
for the $X^{(Z-4)+}$ ions be in part
an artifact of the small size of the system; 
total energies for the isovalent Mg atom (Table~\ref{tab:mg})
are most accurate with 
BD0+tTPSS and BD0+trSCAN and trends are similar to those 
observed for the Ne atom on Table~\ref{tab:neon}, with 
BD0+DFT providing great improvement over BD0.

\begin{table*}
  \begin{center}
    \caption{Accurate~\cite{Petersson1981}
      total energies (in Hartrees) for the 
      Beryllium isoelectronic series and deviations 
      (calculated $-$ exact, in milihartrees) from these values for TPSS, SCAN, BD0, and BD0+DFT using 
      a Cartesian cc-pCVQZ basis. Results from CCD0 and CCD0+DFT are very similar 
      to those from  their BD0 counterparts.}
    \label{tab:beiso}
    \scalebox{0.77}{
      \begin{tabular*}{1.3\textwidth}{@{\extracolsep{\fill}}  lcccccccccc  }
        \hline
                 &    &   &  & &   \multicolumn{6}{c}{BD0+} \\
                 \cline{6-11} 
                 Species & Accurate & TPSS & SCAN & BD0 & pTPSS & tTPSS & pSCAN & tSCAN & prSCAN & trSCAN \\
                 \hline
                 Be        & -14.6674  & -4.2 & 17.3  & 4.0 & -2.7 & -6.0 & 6.0  & -11.1 & -3.7 & -7.5  \\
                 B$^{+}$  & -24.3498  & 3.0  & 31.8  & 6.0 & -2.4 & -6.5 & -6.8 & -13.2 & -4.2 & -9.3  \\
                 C$^{2+}$  & -36.5352  & 8.8  & 45.2  & 6.2 & -3.2 & -7.9 & -8.4 & -15.7 & -5.8 & -11.7 \\
                 N$^{3+}$  & -51.2234  & 15.6 & 59.9  & 7.1 & -3.0 & -8.0 & -8.8 & -16.7 & -6.2 & -12.9 \\
                 O$^{4+}$  & -68.4128  & 22.8 & 75.4  & 8.3 & -2.3 & -7.5 & -8.5 & -17.0 & -6.1 & -13.3 \\
                 F$^{5+}$  & -88.1022  & 29.5 & 90.2  & 8.7 & -2.2 & -7.7 & -8.9 & -17.7 & -6.6 & -14.2 \\
                 Ne$^{6+}$ & -110.2921 & 36.4 & 105.1 & 9.2 & -2.0 & -7.6 & -9.0 & -18.1 & -6.8 & -14.8 \\
        \hline 
         ME                &           & 16.0 & 60.7 & 7.1 & -2.5 & -7.3 & -8.1 & -15.6 & -5.6 & -12.0 \\
         MAE               &           & 17.2 & 60.7 & 7.1 &  2.5 &  7.3 &  8.1 &  15.6 &  5.6 &  12.0 \\
         NPE               &           & 40.6 & 87.9 & 5.2 &  1.2 &  2.0 &  3.0 &   7.1 &  3.1 &   7.3 \\
        \hline
      \end{tabular*}
    }
  \end{center}
\end{table*}

\begin{table*}
  \begin{center}
    \caption{Accurate~\cite{Zhao2008} total
      energy for the Magnesium atom (Hartrees) and deviations 
      (calculated $-$ exact, in milihartrees) from this value for BD0 and BD0+DFT methods using 
      a Cartesian cc-pwCV5Z basis.}
    \label{tab:mg}
    \scalebox{0.8}{
      \begin{tabular*}{1.25\textwidth}{@{\extracolsep{\fill}}  cccccccc  }
        \hline
                 &       &     \multicolumn{6}{c}{BD0+} \\
                 \cline{3-8} 
                 Accurate & BD0 & pTPSS & tTPSS & pSCAN & tSCAN & prSCAN & trSCAN \\
                 \hline
                 -200.053 & 143 & 48 & 1 & 24 & -36 & 38 & -15 \\
        \hline 
      \end{tabular*}
    }
  \end{center}
\end{table*}

\begin{figure}
  \centering
  \includegraphics[width=0.45\textwidth]{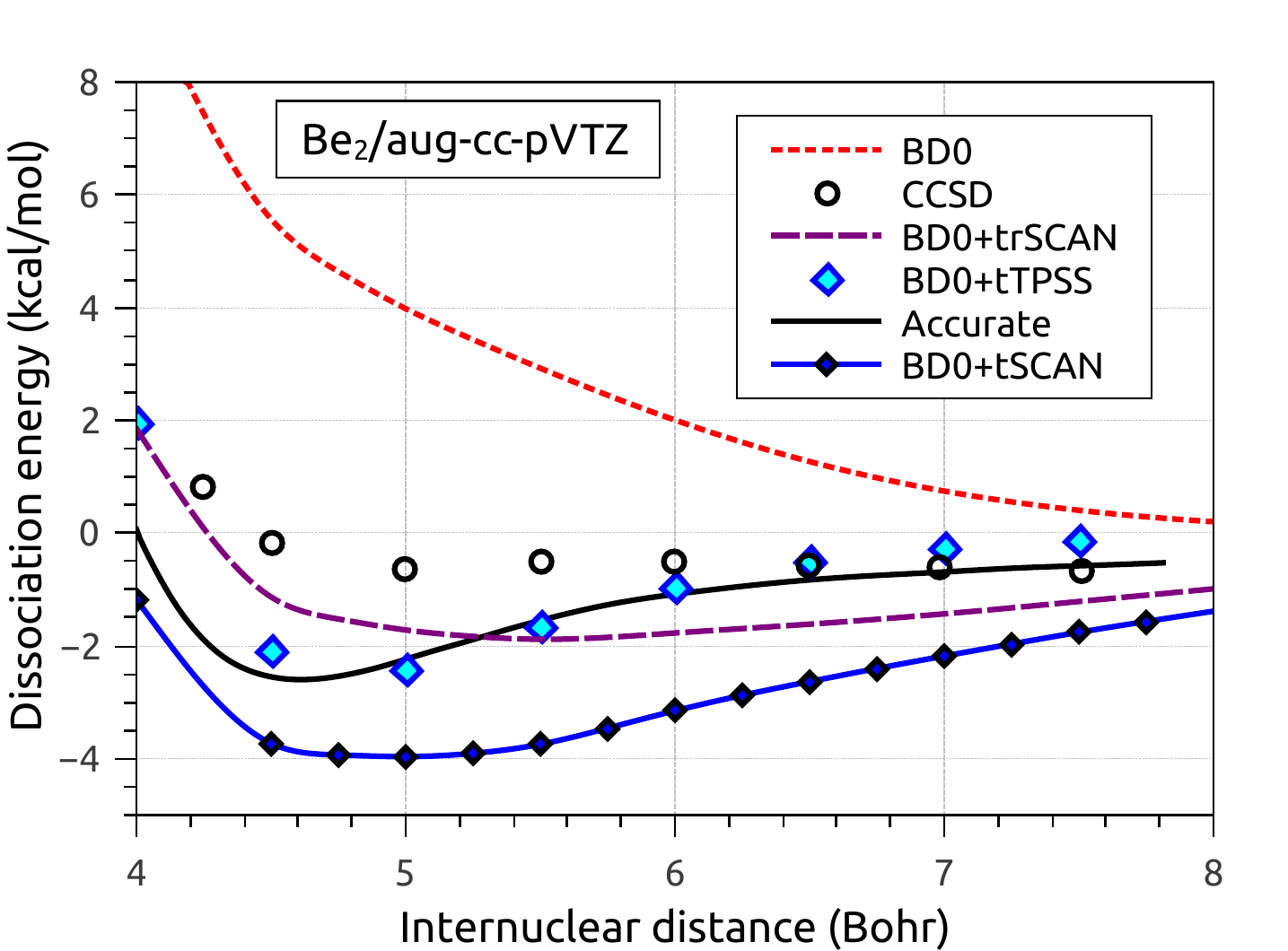}
  \includegraphics[width=0.45\textwidth]{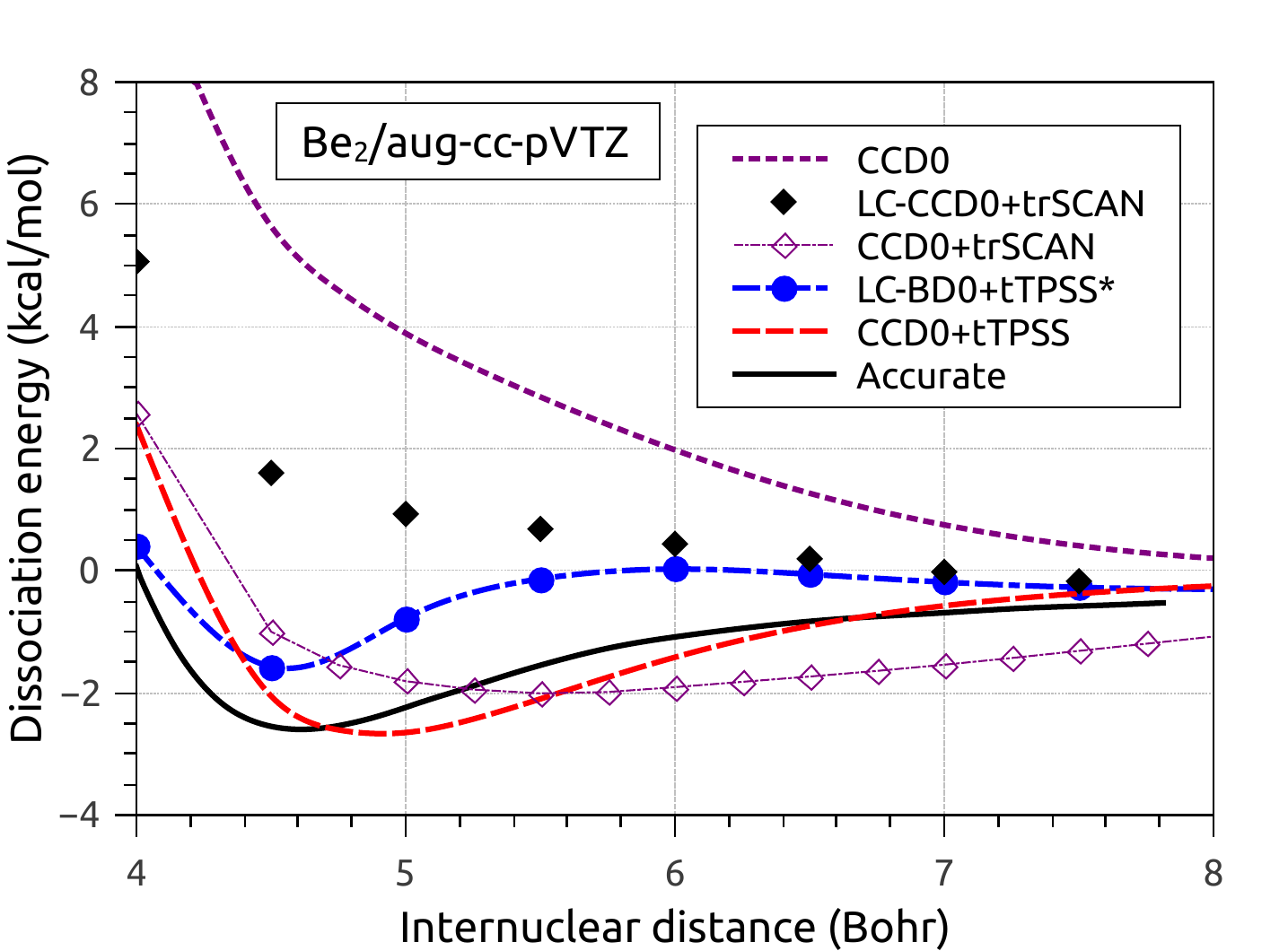}
  \caption{Dissociation energy profiles for 
    the Beryllium dimer calculated by various methods using 
    an uncontracted Cartesian aug-cc-pVTZ basis.
    The accurate data are from explicitly correlated 
  $r_{12}$MRCI calculations from Ref.~\cite{Gdanitz1999}.
    The CCSD data are from Ref.~\cite{Lotrich2005}.
    The curve marked as LC-BD0+tTPSS* uses the 
    semiempirical prefactor of 1.5 on the dRPA correlation 
    of Ref.~\cite{Janesko2009}. }
\label{fig:be2}
\end{figure}

In the tests for static correlation thus far analyzed,
BD0 is able to provide a qualitatively correct description of the 
problem at hand. 
A challenging benchmark for which this is not the case 
is the dissociation of the Beryllium dimer. 
This dissociation is chemically different from the breaking 
of typical bonds discussed above: 
Be$_2$ has a formal (MO-theory) bond order of zero; it is weakly bound
($\approx$ 2 kcal/mol) but not by dispersive 
van der Waals forces, but by a mixture of dynamic and static 
correlation~\cite{Khatib2014}. 
Because of this, common coupled cluster and perturbation methods 
fail for this system; multireference wavefunctions may or may not
work depending on the size of the active space and on how 
dynamic correlation is 
treated~\cite{Khatib2014,Lotrich2005,Gdanitz1999}. 
Figure~\ref{fig:be2} shows that BD0 also fails badly 
for Be$_2$; the potential energy curve is repulsive near 
the accurate equilibrium distance (determined by an explicitly 
correlated MR wavefunction, which agrees with 
experiment~\cite{Lotrich2005,Gdanitz1999}). 
In contrast, BD0+DFT methods bind the Be dimer, with BD0+tTPSS 
giving particularly good qualitative and quantitative agreement 
with the accurate reference data. 
In this case,
BD0- and CCD0-based provide similar results.
LC-BD0+tTPSS affords a good description of the dissociation 
when  the semiempirical prefactor of 1.5 on the dRPA correlation 
of Ref.~\cite{Janesko2009} is used; otherwise the relative energy
at the equilibrium distance may be too high (albeit there is 
still substantial improvement over BD0). 
It has been noted that connected triples are needed to 
obtain sizable bonding~\cite{Sosa1988}; the ability of BD0+DFT 
to do this is thus a good indicative that the way in which the 
methods are mixed is adequate. 
The dissociation of Be$_2$ is an example of the importance 
of the correlation missing in CC0 for obtaining qualitatively 
correct results in difficult cases where a simultaneous, balanced
description of both static and dynamic correlation is necessary.

\section{Conclusions, outlook, and possible improvements}

We have presented techniques to incorporate the correlation that is 
missing in singlet-paired coupled cluster via
density functionals or combinations of these with the dRPA. 
These methods are nonemprical and physically motivated, 
and the benchmarks studied here demonstrate that they are capable 
of describing static and dynamic correlation (including long-range
weak interactions) without symmetry breaking. 
Typical problems of MR+DFT methods such as double counting 
and the symmetry dilemma are avoided in CC0+DFT.
In general, CCD0- and BD0-based
methods provide similar results for weakly correlated systems, 
but BD0+DFT is preferable for problems with static correlation such 
as bond breaking. 
Similarly, both CC0+TPSS and CC0+SCAN yield good overall 
results, although SCAN is better at capturing long-range 
weak interactions. 
Addition of the full triplet-pairing component of $T_2$ to 
CC0 (+tDFT) is frequently better than adding just the 
parallel-spin correlation (+pDFT), in agreement with 
theoretical arguments outlined in Section 2.2.

It is often said of traditional coupled cluster methods that 
they give ``the right answer for the right reason.'' Yet, 
these approximations fail in the presence of strong correlation. 
Here,  LC-CCD0+DFT, for example, can be considered as an attempt 
to avoid this failure while
obtaining the right answer for the right reason via a different 
route: CCD0 correctly describes most of the opposite spin 
correlation without breaking down in the strong correlation regime; 
DFAs are most accurate in the
short-range~\cite{Toulouse2004}, while 
RPA is most accurate in the 
long-range~\cite{Angyan2005,Dobson2006,Yan2000}; 
and we add specifically 
the correlation absent in CCD0 from these last two. 
Nonetheless, there is room for improvement over the techniques 
presented here:
\begin{itemize}
  \item The major drawback of CCD0 and BD0 is their $\mathcal{O}(N^6)$
    scaling, which determines the cost of the combinations 
    presented here. Additionally, the DFA correlation added 
    to CCD0/BD0 is evaluated in terms of a multiple of the
    same-spin correlation. This has the disadvantage that 
    resolutions for the parallel-spin correlation are imperfect, 
    and also that same-spin correlation typically makes up for a larger 
    fraction of the total correlation in
    the uniform electron gas as compared to 
    molecules~\cite{Grafenstein2005,Stoll1978}, which may 
    result in too low total energies in CC0+DFT. 
    Both of these problems can be fixed by appropriate combinations of DFT 
    with pCCD, rather than BD0 and CCD0. 
    The scaling of pCCD is only 
    $\mathcal{O}(N^3)$ (neglecting the
    basis transformation of the two-body interaction~\cite{Henderson2014}).  
    Although pCCD misses more dynamic correlation than CCD0, 
    the intrapair correlation is described almost exactly; 
    pCCD closely reproduces  seniority-zero 
    FCI---an optimal linear combination of all 
    configurations that 
    preserve electron pairs~\cite{Henderson2014} 
    (\textit{i.e.}, all seniority zero configurations; 
    see Ref.~\cite{Bytautas2011} 
    for a detailed explanation of the concept of seniority). 
    Hence, following the philosophy of the present work, 
    we can add the correlation missing in pCCD by switching 
    the discussion from the singlet/triplet-pairing components of CC0 to
    intra/inter-pair correlation. 
    Ways of extracting intra/inter-pair correlation from DFAs 
    have been discussed by 
    Savin \textit{et al.}~\cite{Savin1984}.
    Likewise, adding dRPA correlation to pCCD without double 
    counting can be done with a similar strategy to the one done here
    by including only integrals that ``break'' electron pairs 
    (seniority nonzero) in the $\mathbf{B}$ matrix of dRPA. 

  \item Here, the DFA correlation is added to CC0 in a non-self-consistent
    manner. Although the effect of self-consistency is often small
    when adding dynamic DFA correlation to a 
    wavefunction~\cite{Stoll1985}, 
    it is possible that self-consistency can improve CC0+DFT. 
    In particular, it would be possible to make the singlet-
    and triplet-pairing components of $T_2$ to ``talk'' to 
    each other---and without introducing the risk of breakdown in 
    strongly correlated systems---by including an 
    effective one-body potential from the DFA
    in core Hamiltonian of CC0.     

  \item We have relied here on educated guesses 
    for extracting the parallel spin and short-range 
    correlation from existing functionals. 
    Thus, our CC0+DFT methods
    can be refined by parametrizing a 
    parallel spin correlation functional complementary to
    CC0, and doing a rigorous 
    parametrization of said functional
    for the range separation.
    Related to this, LC-CC0+DFT may also benefit 
    from a finer tuning of the range separation 
    parameter. 

\end{itemize}
Thus, it is likely that the already good results obtained here
can be improved further, and that the cost of CC0+DFT be reduced 
by using pCCD+DFT combinations. 
Indeed, 
many other CC0+DFT (or CC0+RPA) mixtures can be developed following 
the general strategies used here. 
Currently, we are working on some of these
extensions and on testing CC0+DFT methods 
on larger systems; preliminary results have been encouraging.
We hope that the ideas and results presented here 
can stimulate the further development of CC0+DFT methods.

\section*{Acknowledgments}

This work was supported as part of the Center
for the Computational Design of Functional Layered Materials,
an Energy Frontier Research Center funded by the U.S. Department
of Energy, Office of Science, Basic Energy Sciences under
Award \# DE-SC0012575.
GES is a Welch Foundation chair (C-0036).

\end{document}